\documentclass[prb,superscriptaddress,amsfonts,amssymb,amsmath,floats,twocolumn,aps,footinbib,shownopacs]{revtex4-2}
\usepackage{amsmath}
\usepackage{amsfonts}
\usepackage{graphicx}
\usepackage{color}
\usepackage{xcolor}
\usepackage{float}
\usepackage{hyperref}
\usepackage{soul}

\newcommand{\bra}[1]{\langle #1|}
\newcommand{\ket}[1]{|#1\rangle}
\newcommand{\inpro}[2]{\langle #1 | #2 \rangle}
\newcommand{\modif}[1]{{\color{blue} #1}}

\begin{document}
\title {Fermion parity switches imprinted in the photonic field of cavity embedded Kitaev chain}
\author{Victor Fernandez Becerra}
\affiliation{CPHT, CNRS, École polytechnique, Institut Polytechnique de Paris, 91120 Palaiseau, France}
\author{Olesia Dmytruk}
\affiliation{CPHT, CNRS, École polytechnique, Institut Polytechnique de Paris, 91120 Palaiseau, France}

\date{\today}

\begin{abstract}

%\modif{
We study a finite-length Kitaev chain coupled to a single mode photonic cavity. The topological phase of the finite-length Kitaev chain is characterized by the presence of fermion parity switching points that correspond to the degeneracy between even and odd parity ground states. Using exact diagonalization, we compute the many-body energy spectrum of the electron-photon Hamiltonian and we find that the ground state in the topological phase of the Kitaev chain is only weakly affected by the cavity coupling. This is in contrast with the excited states showing strong dependence on the cavity frequency. We find that the photon number and the photonic field quadratures peak at values of the chemical potential corresponding to parity switching points revealing a property of the finite-length Kitaev chain in the topological phase. This later finding suggests that quantum optics experiments could be used to detect topological features of the Kitaev chain embedded into a photonic cavity. Moreover, calculations of photonic quadratures reveal squeezed states that are both captured by the exact diagonalization technique and mean field decoupling. However, the mean field approach fails to correctly capture the photonic probability in the odd photonic states.
%}

\end{abstract}

\maketitle

\section{Introduction}

The strong coupling of light with quantum materials offers new avenues to control their properties and potentially induce new phenomena. For example, experiments with ultrafast laser pulses have shown the anomalous Hall effect in graphene~\cite{McIver_exp_AHE}, Floquet-Bloch bands in a topological insulator~\cite{Wang_Floquet_BiSe}, and the emergence of metastable superconducting phases with enhanced critical temperature~\cite{Mitrano_LightInd_SC,Budden_LightInd_SC}. Embedding materials in photonic cavities serves as another platform for investigating strong light-matter coupling~\cite{CavRevw_Schlawin}. Striking phenomena such as the modification of the Quantum Hall effect~\cite{appugliese2022breakdown} and metal-to-insulator phase transition in a charge density wave material~\cite{Jarc_exp_CDW} have been reported in cavity-embedded systems. 

Topological materials, such as topological superconductors and topological insulators~\cite{TopoMatter_Rev_Hasan, TopoMatter_Rev_Qi, TopoSup_Rev_Sato} might also benefit from the strong coupling to light. Paradigmatic topological models such as the Kitaev chain~\cite{kitaev2001unpaired}, Su-Schrieffer-Heeger chain~\cite{SSH_paper}, and other one-dimensional and ladder models of topological superconductors~\cite{Majo_wires_Lutchyn,Majo_wires_Oreg,CreutzMajorana}  have been embedded into photonic cavities to theoretically investigate the strong coupling to light.  Exotic light-matter states have been revealed, namely Majorana polaritons~\cite{trif2012resonantly, bacciconi2023topological, dmytruk2024hybrid}. Moreover, the control of topological phases with quantum light ~\cite{ciuti2021cavity,dmytruk2022controlling,perez2023light,shaffer2023entanglement,perez2023many,nguyen2023electron,nguyen2024electron,shaw2024theoreticalstudycavitymodulatedtopological}, and the implementation of Majorana parity-based qubits~\cite{Z2Qubit_Yavilberg,trif2019braiding,contamin2021topological} are among other findings reported so far. On the other hand, graphene, albeit not a topological material, is predicted to show a light-induced topological phase when embedded into a cavity~\cite{ChernGraphene_Wang, li2022effective,dag2024,ghorashi2025tunable,tay2025terahertz}. Another electronic system that is not topological but conceptually intriguing is double quantum dots connected to superconducting islands to realize poor man's Majorana bound states (MBS)~\cite{sau2012realizing,leijnse2012parity,fulga2013adaptive,SeoaneSouto2024}, zero-energy modes engineered to resemble the MBS of the Kitaev model but lacking topological protection~\cite{dvir2023realization,zatelli2023robust,tenhaaf2023engineering, bordin2024signatures}. Experimental advances in the embedding of quantum dots into photonic cavities~\cite{CavDQDsuper2018, CavDQDsuper2019} lay the ground for potential implementations of poor man's MBS strongly coupled to quantum light. A benefit from such implementations could arise, such as tuning improvement of the poor man's MBS~\cite{gomez2024high}.

The hybridization between light and matter in cavity-embedded systems also permits us to turn the focus to photonic quantities. Photon occupations have been computed in strongly correlated and topological electronic systems coupled to single-mode cavities~\cite{passetti2023cavity,Li_ED_Hubbard,Hubbard_Cav_Nakamoto,kass2024manybodyphotonblockadequantum,PhotonNumber2020}. Moreover, recent proposals have predicted the possibility of indirectly measuring electronic correlations via photonic experiments~\cite{PhotElec_measure_Lysne, grunwald2024cavity}. Field quadratures are other important photonic quantities that are experimentally accessible~\cite{QDsqueeze_Schulte, optomech_Brooks, optomech_Safavi-Naeini}. Their measurements have become central in the detection of squeezed light, the state in which the variance in one of the field quadratures is smaller than the variance of coherent light or the vacuum~\cite{squeeze_predict, squeezeRev_Andersen}. Squeezed light is conceptually interesting for its behavior as light with no classical counterpart but also for its technological applications in the detection of gravitational waves~\cite{GravWaves2011,GravWaves2013} and its utility for quantum information processing~\cite{squeeze_QComp_Zhong, Squeeze_QComp_Madsen}. 
%%%%%%%%%%%%%%%%%%%%%%%%%%%%%%%%%%%%%%%%%%%%%
%%%%%%%%%%%%%%%%%%%%%%%%%%%%%%%%%%%%%%%%%%%%%
\begin{figure}[htb]
   \centering
   \includegraphics[width=0.49\textwidth]{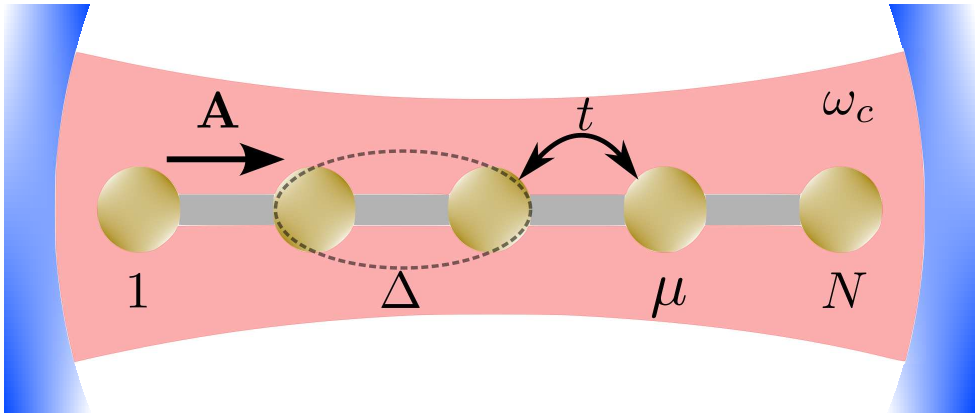} 
    \caption{ Kitaev chain embedded in a single mode photonic cavity with frequency $\omega_c$ and photonic vector potential $\mathbf{A}=(g/e)(a^{\dagger} +a)\mathbf{e}_x$, where $\mathbf{e}_x$ points along the chain and $g$ is the light-matter coupling constant. The Kitaev chain is defined by the hopping $t$, chemical potential $\mu$, superconducting gap $\Delta$ and the number of sites $N$.}
   \label{setup}
\end{figure}
%%%%%%%%%%%%%%%%%%%%%%%%%%%%%%%%%%%%%%%%%%%%%
%%%%%%%%%%%%%%%%%%%%%%%%%%%%%%%%%%%%%%%%%%%%%

In this work, we investigate the light-matter correlations arising in a topological superconductor embedded into a single mode photonic cavity. 
Motivated by the experimental realizations of two and three sites Kitaev chains~\cite{dvir2023realization,zatelli2023robust,tenhaaf2023engineering, bordin2024signatures}, for the topological superconductor we consider the Kitaev chain model and couple it to the cavity via a quantum Peierls substitution~\cite{dmytruk2021gauge,dmytruk2022controlling,perez2023light,passetti2023cavity,Li_ED_Hubbard,dmytruk2024hybrid,nguyen2024electron,Kozin2025,kass2024manybodyphotonblockadequantum,Hubbard_Cav_Nakamoto}, in sharp contrast to previous works that considered a coupling of the capacitive type~\cite{trif2012resonantly,dmytruk2015cavity,PhotonNumber2020}.  Using the exact diagonalization in the many-body electron-photon basis, we get access to the full energy spectrum of the Kitaev chain embedded in a cavity and find signatures of the topological phase of the isolated finite-length Kitaev chain imprinted in the photonic quantities, with previous theoretical works focusing on the Majorana polaritons~\cite{trif2012resonantly,dmytruk2024hybrid} and electronic superradiance~\cite{PhotonNumber2020} in such a setup. We find that the amplitude of the Majorana operators is similar to that of MBS in the isolated Kitaev chain. Moreover, for the short chains considered here, the hybridization between the states localized at the opposite ends of the chain leads to ground state fermion parity switches as a function of the chemical potential. The fermion parity switches in the ground state imprint distinctive features in the photonic field that can be detected in measurements of the photonic occupation and the field quadratures. We also report squeezing of light and the photon probabilities in the ground state and some excited states. 
\section{Model} 
%%
%%%%%%%%%%%%%%%%%%%%%%%%%%%%%%%%%%%%%%%%%%%%%
%%%%%%%%%%%%%%%%%%%%%%%%%%%%%%%%%%%%%%%%%%%%%
\begin{figure}[htb]
	\centering
	\includegraphics[width=0.49\textwidth]{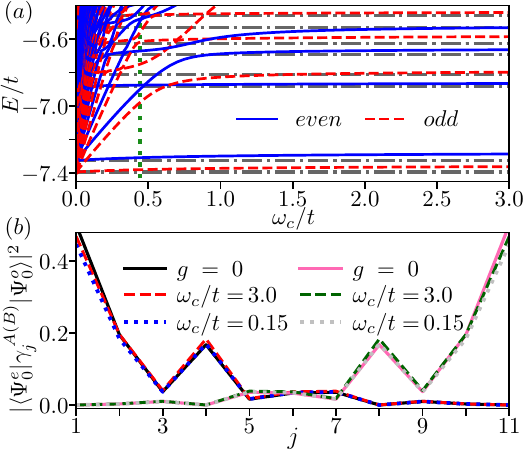}
	\caption{
		(a) Many-body energy spectrum of the Kitaev chain coupled to a single mode cavity as a function of cavity frequency $\omega_c$. The blue solid (red dashed) lines correspond to the even (odd) parity states obtained with ED of Eq.~\eqref{eq:Hc}.
		For comparison, the many-body spectrum of the isolated Kitaev chain ($g=0$) is also shown in gray dot-dashed lines. The green dotted vertical line corresponds to the gap $\Delta_{\text{g}}/t = 0.44$ of the isolated Kitaev chain. (b) Amplitude of the Majorana operators $\gamma_j^A = c^{\dagger}_j +  c_j$ and $\gamma_j^B = i(c^{\dagger}_j - c_j)$ between the even $|\Psi^e_0\rangle$ and odd parity $|\Psi^o_0\rangle$ ground states.  The black solid, red dashed and blue dotted lines correspond to the amplitude $|\bra{\Psi_0^e}\gamma_j^A\ket{\Psi_0^0}|^2$ for the isolated Kitaev chain ($g=0$), and cavity frequencies  $\omega_c/t = 3$ and $\omega_c/t = 0.15$. The pink solid, green dashed and gray dotted lines correspond to the amplitude $|\bra{\Psi_0^e}\gamma_j^B\ket{\Psi_0^0}|^2$ for the isolated Kitaev chain ($g=0$), and cavity frequencies  $\omega_c/t = 3$ and $\omega_c/t = 0.15$.  The localization of the amplitude of the Majorana operators at the edges of the chain shows small changes as a function of frequency. The parameters of the system are: $\mu/t = 0.75$ $\Delta/t = 0.2$, $g=0.25$, $N=11$, and $N_c = 20$. 
 }
	\label{fig:ManyBodySpectrum}
\end{figure}
%%%%%%%%%%%%%%%%%%%%%%%%%%%%%%%%%%%%%%%%%%%%%
%%%%%%%%%%%%%%%%%%%%%%%%%%%%%%%%%%%%%%%%%%%%%
We consider the Kitaev chain coupled to a single mode photonic cavity (see Fig.~\ref{setup}). The Hamiltonian~\cite{kitaev2001unpaired} of a one-dimensional Kitaev chain composed of $N$ sites is given by
\begin{align}
H_K &= -\mu\sum_{j=1}^{N} \left(c_j^\dag c_j - \dfrac{1}{2}\right)- t\sum_{j=1}^{N-1}\left(c_j^\dag c_{j+1} + \text{h.c.}\right) \notag\\
&+\Delta\sum_{j=1}^{N-1}\left(c_j c_{j+1} + \text{h.c.}\right),
\label{eq:KitaevH}
\end{align}
where $c_j^\dag $ and $c_j$ are fermionic creation and annihilation operators at site $j$, respectively. Here, $\mu$ is the chemical potential, $t$ is the hopping amplitude, and $\Delta$ is a $p$-wave superconducting pairing potential. The Kitaev chain hosts two Majorana bound states localized at the opposite ends of the chain in the topological phase when $|\mu| < 2t$.

To couple the Kitaev chain to a single mode photonic cavity with frequency $\omega_c$ described by $H_{ph} = \omega_c a^\dag a$, where $a^\dag$ ($a$) are photonic creation (annihilation) operators, we consider a uniform photonic vector potential parallel to the Kitaev chain, $\mathbf{A} = g/\left(e\sqrt{N}\right)\left(a^\dag + a\right)\mathbf{e}_x$, with $g$ being the light-matter coupling strength. The light-matter coupling Hamiltonian is derived by applying a unitary transformation $U = e^{i \frac{g}{\sqrt{N}} (a+a^\dag)\sum_j (j-j_0) c_{j}^\dag c_{j} }$, with $j_0$ being a constant along the chain, to the electronic Hamiltonian~\cite{dmytruk2021gauge,dmytruk2022controlling, dmytruk2024hybrid}, $H_{c} = U^\dag H_K U\notag + \omega_c a^\dag a$. We arrive at
\begin{align}
    H_{c} &= -\mu\sum_{j=1}^{N}\left(c_j^\dag c_j - \dfrac{1}{2}\right)- \sum_{j=1}^{N-1}\left(t e^{i\frac{g}{\sqrt{N}}(a+a^\dag)} c_j^\dag c_{j+1} + \text{h.c.}\right)\notag\\
&+\sum_{j=1}^{N-1}\left(\Delta e^{i\frac{g}{\sqrt{N}}\phi_j (a+a^\dag)}  c_j c_{j+1}+  \text{h.c.}\right) + \omega_c a^\dag a,
\label{eq:Hc}
\end{align}
where $\phi_j = 2\left(j - j_0+1/2\right)$. To preserve the inversion symmetry between the two ends of the Kitaev chain we choose $\phi_1 = -\phi_{N-1}$~\cite{perez2022topology,dmytruk2024hybrid}. %But it is also interesting to consider how the Majorana probability density is 

\section{Many-body energy spectrum.} 
%%%%%%%%%%%%%%%%%%%%%%%%%%%%%%%%%%%%%%%%%%%%%
%%%%%%%%%%%%%%%%%%%%%%%%%%%%%%%%%%%%%%%%%%%%%
\begin{figure}[ttt]
    \centering
    \includegraphics[width=0.49\textwidth]{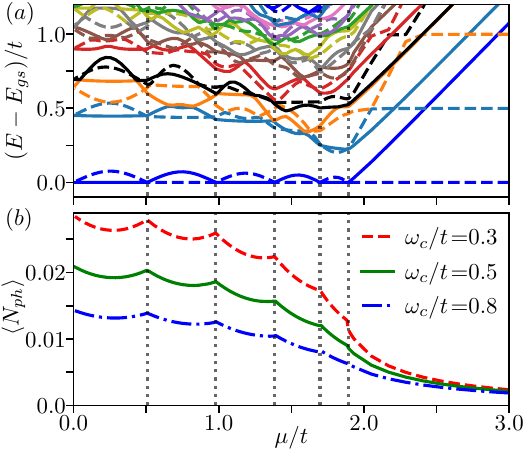}
    \caption{(a) Many-body energy spectrum, ground state energy subtracted, of the cavity-embedded Kitaev chain as a function of the chemical potential $\mu$. States with even (odd) parity are plotted with straight (dashed) lines. For $\mu/t < \mu_{ps}^1$, where $\mu_{ps}^1 = 2\sqrt{t^2-\Delta^2}\cos\left(\pi/(N+1)\right)$, the parity of the ground state alternates between even and odd, and a doubly degenerate ground state takes place at the parity switching points. The cavity frequency considered here is $\omega_c/t = 0.5$ and no considerable changes are found in the behavior of the ground state  within the interval $0.1 < \omega_c/t < 10$. (b) Photon number at the ground state as a function of chemical potential for frequencies indicated in the figure. The photon number shows peaks at the parity switching points. Dotted vertical lines are introduced as a guide to the eye to indicate where the parity switches. The parameters of the system are the same as in Fig. \ref{fig:ManyBodySpectrum}.
}
    \label{fig:PhotonNumber}
\end{figure}
%%%%%%%%%%%%%%%%%%%%%%%%%%%%%%%%%%%%%%%%%%%%%
%%%%%%%%%%%%%%%%%%%%%%%%%%%%%%%%%%%%%%%%%%%%%

In this section, we discuss the many-body energy spectrum of the Kitaev chain coupled to cavity photons.  The energy spectrum of Hamiltonian~\eqref{eq:Hc} is computed using exact diagonalization (see Appendix~\ref{solvers}). Exact diagonalization (ED) allows one to obtain all the states 
and is valid for any value of the system parameters, such as the light-matter coupling strength $g$ and cavity frequency $\omega_c$, with the only limitation coming from the number of sites in the chain. In finite-size chains and in the absence of the photon coupling ($g=0$) the many-body spectrum of the isolated Kitaev Hamiltonian contains $2^{N}$ energy levels and two distinct phases, the trivial and non-trivial topological phases~\cite{kitaev2001unpaired}. The ground state in the non-trivial phase contains two levels with an energy splitting proportional to $e^{-N/N_0}$, with $N_0$ a constant defined in terms of physical parameters~\cite{kitaev2001unpaired}. In long chains ($N \gg N_0$) the ground state is degenerate. On the other hand, in short chains the ground state is degenerate only at the parity switching points~\cite{hegde2015quench, hegde2016majorana,rancic2022}.

In what follows, we consider the Kitaev chain with $N=11$ lattice sites and study the behavior of the many-body energy spectrum as a function of cavity frequency. The cutoff in the number of photonic Fock states in our ED algorithm is $N_c = 20$. 
With the exception of frequencies near zero, Fig.~\ref{fig:ManyBodySpectrum}(a) shows that the ground state is split into two levels with even and odd parities, namely $E_0^e$ and $E_0^o$, respectively. This splitting of the ground state is similar to the splitting in the isolated Kitaev chain, shown in the figure with gray dot-dashed lines. Notice that the energy levels of the excited states $E_{l>0}^{e(o)}$ change considerably with the cavity frequency, especially at low frequencies; therefore, it is convenient to introduce the band gap and bandwith of the isolated Kitaev chain to analyze the energy spectrum of the cavity embedded system. The band gap of the isolated Kitaev chain is the minimum energy that separates the ground state from the excited states, namely $\Delta_g$. In the figure $\Delta_g$ is plotted with a green dotted line. The bandwith of the isolated Kitaev chain is defined as $E_{bw} = E_{mx}-E_{gs}$, with $E_{gs}$ ($E_{mx}$) the energy of the ground state (the highest excited state). For $\omega_c\ll\Delta_g$ the energy levels 
lie within the interval $[E_{gs},E_{gs}+E_{bw}]$, leaving the many-body energy spectrum of the Kitaev chain coupled to photons gapless. With increasing $\omega_c$ the energy levels of the excited states now grow linearly with $\omega_c$ and a band gap opens up in the cavity embedded system. Moreover, we also start noticing in the excited states characteristic anticrossings, or energy level repulsion as is also know~\cite{LevRepulsion2025, LevelRepulsion2016}, due to the strong hybridization between light and matter~\cite{Anticross_Reithmaier, Anticross_Yoshie, Anticross_Peter}. Finally, when $\omega_c > \Delta_g$ the energy levels no longer grow linearly with $\omega_c$ and their energies approach those of the isolated Kitaev chain. This could be analytically shown with high frequency expansion, see Appendix~\ref{solvers}. More details on the many-body energy spectrum of Eq.~\eqref{eq:Hc} for a shorter chain are discussed in Appendix~\ref{PhotoBands}. \\

 Introducing the Majorana operators $\gamma_j^A = c^{\dagger}_j +  c_j$ and  $\gamma_j^B =  i(c^{\dagger}_j - c_j)$, we compute amplitudes for the ground state of the cavity-embedded system to reveal local electronic properties~\modif{\cite{PhysRevB.93.075129,PhysRevB.111.075170}}. In Fig.~\ref{fig:ManyBodySpectrum}~(b), the amplitudes $|\langle \Psi^e_0| \gamma^A_j|\Psi^o_0\rangle|^2$ and $|\langle \Psi^e_0| \gamma_j^B|\Psi^o_0\rangle|^2$, where $|\Psi^e_0\rangle$ and $|\Psi^o_0\rangle$ correspond to the even and odd parity ground states, show localization at the edges of the chain similar to the localization of MBS in the isolated Kitaev chain. Moreover, the amplitudes increase slightly with increasing $\omega_c$ (see Appendix~\ref{majo_dens_wc} for more details), 
 consistent with a similar behavior in the plot of energy vs. frequency in the ground state (see Fig.~\ref{fig:ManyBodySpectrum}~(a)). This frequency dependence is a manifestation of the hybridization between light and MBS~\cite{bacciconi2023topological}.

 \section{Parity switches}
 
 In this section, we demonstrate that the many-body energy spectrum of the Kitaev chain embedded in a cavity has different features in the topological and trivial phases. The many-body spectrum of Hamiltonian~\ref{eq:Hc} is plotted in Fig.~\ref{fig:PhotonNumber}(a) as a function of the chemical potential for $g=0.25$ and $\omega_c/t = 0.5$. For the sake of visualization we have subtracted from all the energies the energy of the ground state. In isolated short-length Kitaev chains and Majorana billiards the values at which the ground state is degenerate have already been reported~\cite{hegde2015quench,hegde2016majorana,rancic2022, Majobilliards}. The crossings or parity switching points are given by the formula $\mu_{ps} = 2\sqrt{t^2-\Delta^2}\cos\left(p\,\pi/(N+1)\right)$, where $p = 1,2, . . . ,M/2$, with $M=N$ $(M=N-1)$ for even (odd) $N$. In the figure, the values of $\mu_{ps}$ are plotted with gray dotted lines.  Notice that while the parity of the ground state alternates between even and odd for $|\mu| < \mu_{ps}^1$, with $\mu_{ps}^1 = 2\sqrt{t^2-\Delta^2}\cos\left(\pi/(N+1)\right)$, it remains odd for $|\mu| > \mu_{ps}^1$. Notice also for $|\mu| < \mu_{ps}^1$ the existence of a doubly degenerate ground state at the parity switching points, a consequence of MBS hosted in the system and strongly hybridized due to the finite length of the chain and the light-matter coupling. The formula for $\mu_{ps}$ in the isolated Kitaev chain provides an approximation of the parity switching points in the Kitaev chain coupled to cavity photons (Eq.~\ref{eq:Hc}). In Appendix~\ref{phasediag} we plot the dependence of one parity switching point on the light-matter coupling strength and compare it with its equivalent in the isolated Kitaev chain. Given the small deviation between the two values compared, the formula for $\mu_{ps}$ in the isolated Kitaev chain will be employed in the rest of the article as a reference.

\section{Photonic properties} 
%%%%%%%%%%%%%%%%%%%%%%%%%%%%%%%%%%%%%%%%%%%%%
%%%%%%%%%%%%%%%%%%%%%%%%%%%%%%%%%%%%%%%%%%%%%
\begin{figure}[ttt]
    \centering
    \includegraphics[width=0.48\textwidth]{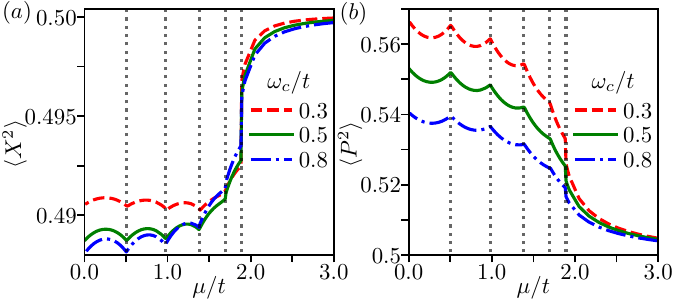}
    \caption{Photonic quadratures (a) $\langle X^2\rangle$ and (b) $\langle P^2\rangle$, with $X=(a+a^{\dag})/\sqrt{2}$ and $P=-i(a-a^{\dag})/\sqrt{2}$, computed at the ground state as a function of chemical potential $\mu/t$ for different cavity frequencies $\omega_c/t$, indicated directly in the plot. While the values of $\langle P^2\rangle$ decrease with increasing $\omega_c$ independently of $\mu$, $\langle X^2\rangle$ shows a richer behavior with respect to $\mu$ and $\omega_c$.
    Importantly, the quadratures also show particular features at the parity switching points $\mu_{ps}$, dips in $\langle X^2\rangle$ and peaks in $\langle P^2\rangle$. For $\mu/t \gtrsim 2$ corresponding to the ground state with a well defined parity, the values of the quadratures approach $1/2$ as in the uncoupled case. Dotted vertical lines are introduced as a guide to the eye to indicate where the parity switches. The parameters of the system are the same as in Fig. \ref{fig:ManyBodySpectrum}. 
    }
    \label{Quad:ground}
\end{figure}
%%%%%%%%%%%%%%%%%%%%%%%%%%%%%%%%%%%%%%%%%%%%%
%%%%%%%%%%%%%%%%%%%%%%%%%%%%%%%%%%%%%%%%%%%%%

In this section, we turn our attention to photonic quantities that can be measured directly or indirectly in optical experiments~\cite{CavExp_Bertet, CavExp_Schuster, QDsqueeze_Schulte, optomech_Brooks, optomech_Safavi-Naeini}. Introducing the photon number operator $N_{ph} = a^\dag a$, in Fig.~\ref{fig:PhotonNumber}~(b) we compute the photon number $\langle N_{ph}\rangle$ in the ground state of the electron-photon Hamiltonian~\ref{eq:Hc} at different cavity frequencies. With the guide of the dotted lines indicating the values of $\mu_{ps}$, we reveal peaks in the photon occupation number at the parity crossings of the ground state. Moreover, the positions of the peaks barely change with the cavity frequency, in agreement with the weak energy dependence of the MBS with frequency shown in Fig.~\ref{fig:ManyBodySpectrum}~(a). We note that the photon number computed with mean field decoupling (MFD) (see Appendix~\ref{solvers}) reveals qualitatively the same features as in Fig.~\ref{fig:PhotonNumber}~(b), 
although the magnitudes considerably differ between MFD and ED. The dependence of the photon number on the parity switching points of the Kitaev chain can be understood from the mean field Hamiltonian that demonstrates that $\langle a^\dag a\rangle$ is proportional to the  expectation value of an electronic operator. Expanding the Hamiltonian to second order in the light-matter coupling strength $g$ and taking a large frequency limit, we obtain the relation between the photon number and the expectation value of the diamagnetic current  operator $\langle J_d\rangle$, $\langle a^\dag a\rangle \approx \langle J_d\rangle^2/ \left(4 \omega_c^2\right)$ (see Appendix~\ref{photoN_MF} for the details of the derivation.)

With ED we find that the behavior of the photon number in the excited states is distinct from that of the ground state; see Appendix~\ref{PhotoNExc}. The sharp transitions seen in the photon number at a given excited state coincide with the approach of higher excited states to the said state in the many-body energy spectrum. %\modif{Finally, to justify the number of photonic Fock states chosen in the ED method used, in Appendix~\ref{photonCutoff} we show that $N_c = 20$ safely guarantees convergence of the photon number and the peaks at the parity switching points.}   
Finally, we find that the photon number is independent of the photonic cutoff for $N_c \geq 5$ (see Appendix~\ref{photonCutoff}).

%%%%%%%%%%%%%%%%%%%%%%%%%%%%%%%%%%%%%%%%%%%%%
%%%%%%%%%%%%%%%%%%%%%%%%%%%%%%%%%%%%%%%%%%%%%
\begin{figure}{}
	\centering
	\includegraphics[width=0.48\textwidth]{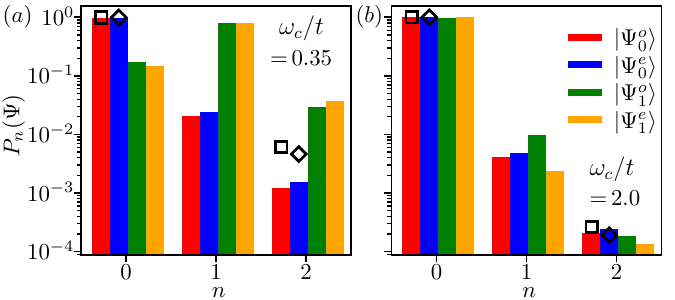}
	\caption{ Probability $P_n(\Psi)=|\bra{n} \Psi \rangle|^2$ to find the light-matter state $\ket{\Psi}$ in the photon state $\ket{n}$ at frequencies $\omega_c/t= 0.35$ (a) and  $\omega_c/t = 2.0$ (b). The light-matter states considered are: the even and odd parity ground state ($\ket{\Psi_0^e}$ and $\ket{\Psi_0^o}$) and the excited states $\ket{\Psi_1^e}$ and $\ket{\Psi_1^o}$. Notice that the probabilities $P_{n=0}(\Psi_0^e)$ and $P_{n=0}(\Psi_0^o)$ are approximately one and weakly dependent on frequency. On the other hand, the probabilities $P_{n=0}(\Psi_1^{e(o)})$, $P_{n=1}(\Psi_1^{e(o)})$, and $P_{n=2}(\Psi_1^{e(o)})$ change considerably with frequency. The photon probabilities computed with MFD are also shown in the figure with empty diamonds (squares) for the even (odd) parity ground state. The parameters of the system are the same as in Fig. \ref{fig:ManyBodySpectrum}.
    }
	\label{fig:PhotonProb}
\end{figure}
%%%%%%%%%%%%%%%%%%%%%%%%%%%%%%%%%%%%%%%%%%%%%
%%%%%%%%%%%%%%%%%%%%%%%%%%%%%%%%%%%%%%%%%%%%%

The next set of quantities we report are related to the photonic quadratures, defined in terms of the operators $X=\left(a+a^{\dag}\right)/\sqrt{2}$ and $P=-i\left(a-a^{\dag}\right)/\sqrt{2}$, usually known as the photonic position and momentum operators, respectively. Gauge invariance imposes $\langle X \rangle = \langle P \rangle = 0$~\cite{Andolina_azero}, then the direct computation of $\langle X^2\rangle$ and $\langle P^2\rangle$ produces the variances of the photonic position and momentum operators, respectively. In Fig.~\ref{Quad:ground} we plot  $\langle X^2\rangle$ (a) and $\langle P^2\rangle$ (b) in the ground state as a function of $\mu$ for different cavity frequencies.  For the parity alternating ground state we observe that the variance of the $X$ quadrature is below the threshold limit of coherent light or the vacuum ($\langle X^2 \rangle - \langle X \rangle^2 < 1/2$), i.e. light is squeezed~\cite{squeeze_predict}. Notice also that while the squeezed quadrature shows dips at the parity switching points $\mu_{ps}$, $\langle P^2\rangle$ shows peaks such as those observed in the photon number of Fig.~\ref{fig:PhotonNumber}(b). The frequency dependence of both quadratures is remarkably distinct, with $\langle P^2\rangle$ monotonically decreasing with $\omega_c$ and $\langle X^2\rangle$ featuring a local minimum whose position and minimum depend on $\mu$ (see Appendix~\ref{Quad}). To guide the reader, the crossing points of the isolated Kitaev chain are also added to the figure with dotted lines. We also computed the variances of the quadratures in the ground state but with MFD (see Appendix~\ref{Quad}). Although the squeezed quadrature obtained with ED and MFD reveal some differences, in the other quadrature, ED and MFD yield qualitatively the same features. Squeezing of the photon can be inferred from the expanded mean field photonic Hamiltonian (see Appendix~\ref{photoN_MF})
\begin{align}
    H_{c,ph}^{mf} = \omega_c a^\dag a - \dfrac{1}{2} (a+a^\dag)^2 \langle J_d\rangle,
    \label{eq:HphMFexpandedMT}
\end{align}
where $\langle J_d\rangle$ is the expectation value of the diamagnetic current operator and is proportional to the squeezing parameter~\cite{dmytruk2022controlling}. 

The probability to find the cavity-embedded Kitaev chain in the photonic state $\ket{n}$ is given by $P_n(\Psi) = |\bra{n} \Psi\rangle|^2$, where $\ket{\Psi}$ is an eigenstate of Hamiltonian~\ref{eq:Hc}. The advantage that ED provides %regarding the calculation of excited states 
allow us to plot in Fig.~\ref{fig:PhotonProb} not only the photon probabilities in the ground state, but also in the excited states $\ket{\Psi^{e(o)}_{l>1}}$. The ground states $\ket{\Psi_0^e}$ and $\ket{\Psi_0^o}$, with even and odd parities, respectively, have probabilities $P_{n=0}(\Psi_0^e) \approx 1$ and $P_{n=0}(\Psi_0^o) \approx 1$ weakly dependent on $\omega_c$ (see the complete frequency dependence in Appendix~\ref{PhProba}). On the other hand, the excited states $\ket{\Psi_1^e}$ and $\ket{\Psi_1^o}$ show a strong dependence with $\omega_c$ in the probabilities $P_{n=0}(\Psi_1^{e(o)})$ and $P_{n=1}(\Psi_1^{e(o)})$. While the dominant probability at $\omega_c/t =0.35$ is $P_{n=1}(\Psi_1^{e(o)})$, at $\omega_c/t =2.0$ the dominant probability is $P_{n=0}(\Psi_1^{e(o)})$. The dependence of the excited states with frequency, especially at low frequencies, is a consequence of the strong light-matter interaction already suggested by the anticrossings of levels $E_1^e$ and $E_1^o$ in the energy spectrum of Fig.~\ref{fig:ManyBodySpectrum}.  The photon probabilities computed with MFD are also shown in the figure with empty squares and diamonds. The probabilities obtained with ED and MFD qualitatively agree, except in $P_{n=1}(\Psi_0^{e(o)})$, where the latter predicts zero probability (not shown in the figure). To explain the disappearance of the photonic probabilities at odd photon numbers, we refer the reader to the expanded mean field photonic Hamiltonian~\eqref{eq:HphMFexpandedMT} that is quadratic in the photonic operators and leads to zero probability to find the photon in a state with odd $n$. A perturbation treatment of the Hubbard model coupled to cavity photons also reported zero photon probabilities in odd photon numbers~\cite{passetti2023cavity,Hubbard_Cav_Nakamoto}.

\section{Conclusions}

In this work, we investigated a short-length Kitaev chain coupled to cavity photons with ED. We analyzed the many-body energy spectrum of the cavity-embedded system and revealed not only features of the ground state but also of the excited states. As a function of chemical potential, we found two distinct types of ground states, one with a well-defined parity and another with an alternating parity stemming from MBS hybridization. The reported MBS are hybridized mainly due to the the short length of the chains considered here. This hybridization leads to fermion parity switches imprinting distinctive features in the photon number and the photonic quadratures. We therefore predict the possibility of detecting the fermion parity switches in quantum optical experiments. We also find quadrature squeezing and report not only the photon number in the 
ground state but also in few excited states.

\section*{acknowledgements}
%{\it Acknowledgments.} 
We would like to thank Mircea Trif, Marco Schirò and Martin Eckstein for useful discussions. 
This work is supported by ERC grant (Q-Light-Topo, Grant Agreement No. 101116525).
%\end{acknowledgements}

\appendix

\section{Formalisms to solve electron-photon Hamiltonians}\label{solvers}
In this section, we discuss several formalisms to solve the Kitaev chain embedded into a single mode photonic cavity. 

\subsection{Exact diagonalization}

To diagonalize Hamiltonian~\eqref{eq:Hc} in the main text we consider product states of type $\ket{\psi}\otimes\ket{\phi}$, where $\ket{\psi}$ ($\ket{\phi}$) is an electronic (photonic) Fock state. The photonic Fock states are generated from the recurrence relation $a^{\dag}\ket{n}= \sqrt{n+1}\ket{n+1}$, where $\ket{n}$ is a state with $n$ photons. The recurrence procedure to create photons is unbounded then the number of photonic Fock states is infinite. To solve the problem numerically a cutoff $N_c$ in the  photonic Fock states has to be set. The Fock space for fermions is finite although it grows exponentially. Moreover, an ordering must be kept to avoid violating the anti-commutation rules. In this work, the creation operators are ordered from left to right according to increasing lattice sites, i.e. $\ket{\psi} = \Pi_{\alpha \in  \{j_1 < j_2 < j_3 < \cdots\}} \,c^{\dag}_{\alpha}  \ket{0}$, where $j_i$ denotes a lattice site. 

The projection of the light-matter Hamiltonian~\eqref{eq:Hc}  into the photonic states $m$ and $n$ reads
\begin{gather}
    \bra{m} H_{c} \ket{n} = \Bigl(m\omega_c  -\mu\sum_{j=1}^{N}\Bigl(c_j^{\dag}c_j -\frac{1}{2}\Bigr)\Bigr)\delta_n^m \notag\\
    - \sum_{j=1}^{N-1}t e^{\frac{-g^2}{2N}}J_{m,n}\Bigl(\frac{g}{\sqrt{N}}\Bigr)
    \left( i^{|m-n|} c_j^\dag c_{j+1} + \text{h.c.}\right)\notag\\
    +\sum_{j=1}^{N-1}\Delta e^{\frac{-g^2\phi_j^2}{2N}}J_{m,n}\Bigl(\frac{g\phi_j}{\sqrt{N}}\Bigr)
    \left( i^{|m-n|} c_j c_{j+1}+  \text{h.c.}\right),
    \label{eq:mn_block}
\end{gather}
where $J_{m,n}$ is a polynomial of order $m+n$ given by
\begin{gather}
    J_{m,n} (x) =
    \begin{cases}
    \sum_{k=0}^n (-1)^k\frac{\sqrt{m!}\sqrt{n!}}{k!(k+m-n)!} \frac{x^{2k+m-n}}{(n-k)!}
    & \text{for  $\;m \geq n$,} \\
    \sum_{k=0}^m (-1)^k \frac{\sqrt{m!}\sqrt{n!}}{k!(k+n-m)!} \frac{x^{2k+n-m}}{(m-k)!}
    & \text{for  $\;m < n$.}
    \end{cases}
    \label{eq:PhotonAmplitude}
\end{gather}
Fermion particle number is not conserved neither in the isolated Kitaev chain, nor in the Kitaev chain coupled to photons. In our diagonalization scheme we then consider all possible particles numbers, from $0,1,\dots N$, where $N$ is the number of lattice sites. The parity operator defined by $\modif{\mathcal{P}} = \Pi_j (1-2c_j^{\dag}c_j)$ has two eigenvalues and commutes with the Hamiltonian (2) in the main text, $[H_c,\modif{\mathcal{P}}]=0$. This conservation of parity allow us to split the Hamiltonian into two blocks, the even $H_c^e$ and $H_c^o$ odd parity blocks,
\begin{align}
    H_c = \left(
    \begin{array}{cc}
    H_c^e & \\
     & H_c^o
    \end{array}
    \right),
\end{align}
with both blocks having dimension $2^{N-1}(N_c+1)\times2^{N-1}(N_c+1)$. The conservation of parity also allows us to label the eigenvalues and eigenstates of Hamiltonian (2) in the main text, e.g. $\ket{\Psi^{e(o)}_l}$ is an eigenstate of block $H_c^{e(o)}$ with energy $E_l^{e(o)}$. Moreover, we also sort the energies within each block in increasing order. 

\subsection{Mean field decoupling of light and matter}
In this section, we use the mean field decoupling (MFD) between electrons and photons to solve Eq.~\eqref{eq:Hc} in the main text. This approach is generally valid in the thermodynamic limit and has been widely used to address the problem of  electronic superradiance~\cite{Andolina_no-go_theorem, dmytruk2021gauge}. However, several recent works have demonstrated that it misses effects caused by light-matter entanglement in finite-length electronic chains coupled to photons~\cite{passetti2023cavity,nguyen2024electron}. 
Writing down the ground state wavefunction of Eq.~\eqref{eq:Hc} in the form $|\Psi\rangle = |\psi\rangle |\phi\rangle$,
where $ |\psi\rangle$ corresponds to the electronic part of the wavefunction and $|\phi\rangle$ stands for the photonic part, we arrive at the  electronic mean-field Hamiltonian with 
\begin{align}
H_{c,el}^{mf} = \langle \phi | H_{c} | \phi \rangle  = -\mu\sum_{j=1}^{N} \left(c_j^\dag c_j - \dfrac{1}{2}\right) \notag\\
- \sum_{j=1}^{N-1}\left(\tilde{t} c_j^\dag c_{j+1} + \text{h.c.}\right)
+\sum_{j=1}^{N-1}\left(\tilde{\Delta}  c_j c_{j+1} +  \text{h.c.}\right),
\label{eq:HelMF}
\end{align}
with $\tilde{t} = t \langle \phi | e^{i\frac{g}{\sqrt{N}}(a+a^\dag)}| \phi \rangle $ and $\tilde{\Delta}_j= \Delta \langle \phi | e^{i\frac{g}{\sqrt{N}}\phi_j (a+a^\dag)}| \phi \rangle $,
and  the photonic mean-field Hamiltonian

\begin{align}
H_{c,ph}^{mf} = \langle \psi | H_{c} | \psi \rangle = \omega_c a^\dag a 
- \left(t_{h} e^{i\frac{g}{\sqrt{N}}(a+a^\dag)}  + \text{h.c.}\right) \notag\\
+\sum_{j=1}^{N-1}\left(\delta_j e^{i\frac{g}{\sqrt{N}}\phi_j (a+a^\dag)}  +  \text{h.c.}\right),
\label{eq:HphMF}
\end{align}
with $t_h = t\langle \psi | \sum_{j=1}^{N-1} c_j^\dag c_{j+1} | \psi \rangle$ and $\delta_j = \Delta\langle \psi |  c_j c_{j+1} | \psi \rangle$. We note that \eqref{eq:HelMF} contains renormalized hopping and pairing amplitudes compared to the isolated Kitaev chain Hamiltonian~(1) in the main text, while a single mode cavity Hamiltonian $H_{\text{ph}}$ becomes strongly modified due to the coupling with the Kitaev chain even at the mean field level~\eqref{eq:HphMF}.
We solve both Eqs.~\eqref{eq:HelMF} and \eqref{eq:HphMF}  self-consistently using the exact diagonalization (in a single particle picture for the electronic mean-field Hamiltonian).

\subsection{High frequency expansion}

Considering that $\omega_c$ is by far the largest of the energy scales, i.e. $\omega_c \gg t, \Delta, \mu$, one can tackle Hamiltonian~\eqref{eq:Hc} in the main text by developing an effective theory based on the Brillouin-Wigner expansion valid in the high frequency limit \cite{HighFrequencyExp,Li_ED_Hubbard,li2022effective}. The effective Hamiltonian reads 
\begin{equation}
	H_{c}^{\text{eff}, \, n} = \bra{n}H_{c}\ket{n} -\sum_{l\neq n} 
	\frac{\bra{n}H_{c}\ket{l}\bra{l}H_{c}\ket{n}}{(l-n)\omega_c}.
	\label{eq:BWapprox}
\end{equation}
For the zero-photon state ($n=0$) we derive an effective Hamiltonian $H_{c}^{\text{eff}, \, 0} = H_{\text{eff}}^{(0)} + H_{\text{eff}}^{(1)}$ using mean-field decoupling $\sum_j O_j\sum_i O_i = \langle \sum_j O_j\rangle \sum_i O_i + \langle \sum_i O_i\rangle \sum_j O_j$.

The zero order term reads
\begin{align}
H_{\text{eff}}^{(0)} &= -\mu\sum_{j=1}^{N}\left(c_j^\dag c_j - \dfrac{1}{2}\right)- \sum_{j=1}^{N-1}\left(t e^{-\frac{g^2}{2N}} c_j^\dag c_{j+1} + \text{h.c.}\right)\notag\\
&+\sum_{j=1}^{N-1}\left(\Delta e^{-\frac{g^2}{2 N}\phi_j^2}  c_j c_{j+1}+  \text{h.c.}\right),
\label{eq:HeffZeroOrder}
\end{align} 
where we note that cavity coupling results in suppression of the hopping amplitude and the superconducting pairing potential.

For the first order term we obtain
\begin{align}
    H_{\text{eff}}^{(1)} &= \sum_{l\neq 0} 
	\frac{1}{l\omega_c}\sum_{j=1}^{N-1}\left(t_{\text{eff}}(l)  c_j^\dag c_{j+1} + \text{h.c.}\right) \notag\\
    &+\sum_{j=1}^{N-1}\left(\Delta_{\text{eff}}(j,l)  c_j c_{j+1}+  \text{h.c.}\right),
\end{align}
where we introduced the effective hopping and superconducting pairing as follows
\begin{align}
t_{\text{eff}}(l) &= -2g_{0,l}^2 t^2 h\left( 1+(-1)^l  \right) \notag\\
&+ 2 g_{0,l}t\Delta \sum_{i=1}^{N-1} \phi_{0,l}(i) p(i)\left(1 +(-1)^l \right),\label{eq:EffectiveHopping}\\
\Delta_{\text{eff}}(j,l) &= -2 \Delta^2\phi_{0,l}(j)  \sum_{i=1}^{N-1}  \Big\{\phi_{0,l}(i)p(i) \left( 1+ (-1)^l\right) \notag\\
&+2t\Delta g_{0,l} \phi_{0,l}(j)h\left(1+(-1)^l \right)\Big\}.
\label{eq:EffectivePairing}
\end{align}	
Here, $g_{0,l} = \langle 0|e^{i\frac{g}{\sqrt{N}}(a+a^\dag)}|l\rangle$ and $
\phi_{0,l}(j) = \langle 0|e^{i\frac{g}{\sqrt{N}}\phi_j (a+a^\dag)} |l\rangle$ can be computed using Eq.~\eqref{eq:PhotonAmplitude}, and $h = \langle \sum_{j=1}^{N-1} c_j^\dag c_{j+1} \rangle$, $p(j) = \langle c_j c_{j+1}\rangle$. 
From Eq.~\eqref{eq:EffectiveHopping} and \eqref{eq:EffectivePairing} we find that one-photon transitions do not contribute to the first order effective Hamiltonian, $H_{\text{eff}}^{(1)}= 0$ for $l = 1,3,5,...$. Since the amplitude of two-photon transitions is much smaller than that of one-photon transitions~\cite{perez2023light}, the leading contribution in the high frequency regime comes from the zero order effective Hamiltonian \eqref{eq:HeffZeroOrder}.

\section{More about the many-body energy spectrum }\label{PhotoBands}

\begin{figure}[htb]
    \centering
    \includegraphics[width=0.49\textwidth]{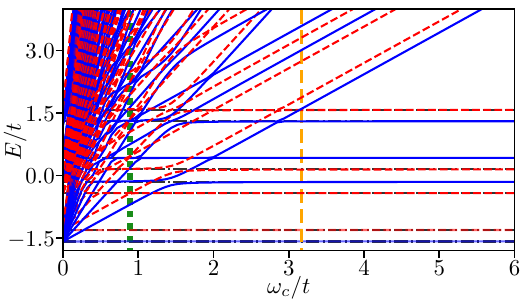}
    \caption{Many-body energy spectrum of the Kitaev chain coupled to photons of a single mode cavity with frequency $\omega_c$. The blue solid (red dashed) lines correspond to the even (odd) parity states obtained with ED of Eq.~(2) in main text. The many-body spectrum of the isolated Kitaev chain (see Eq.~(1) in main text) is also shown in black dash-dotted lines. The green dotted (orange dashed) line corresponds to the band gap (bandwidth) of the isolated Kitaev chain. The parameters of the system are: $N=3$, $\mu/t = 0.3$, $\Delta/t = 0.2$, $g=0.15$, and $N_c = 20$. }
    \label{fig:PhotoBands}
\end{figure}

In this section, we extend the analysis of the energy spectrum as a function of cavity frequency for the many-body Hamiltonian of Eq.~\eqref{eq:Hc}. The energies are calculated via ED for the same parameters as in Fig.~\ref{fig:ManyBodySpectrum}, except for the number of lattice sites, now $N=3$, and the chemical potential, now $\mu/t=0.3$. Given the small number of sites, one can compute with almost zero computational effort all the energies of the system, composed of $N_c+1$ blocks corresponding to the number of photons $n = 0,1,2,..$, with each of the $2^N$ eigenvalues within the block $n$ being shifted with respect to the corresponding eigenvalue of the block $n+1$ by an energy of order $\omega_c$. The full energy spectrum is plotted in Fig.~\ref{fig:PhotoBands}. As discussed in the main text, the energy spectrum shows three different regimes: 1) at $\omega_c \approx 0$ the Kitaev chain coupled to photons becomes gapless, 2) for  $\omega_c \approx \Delta_g$ a gap growing linearly with $\omega_c$ opens up, and 3) for $\omega_c \gg \Delta_g$ the gap becomes $\omega_c$-independent. To discuss more features of the excited states and at higher frequencies we recall the other important energy scale introduced in the main text, the energy bandwidth of the isolated Kitaev chain,  $E_{bw} = E_{mx}-E_{gs}$, with $E_{gs}$ ($E_{mx}$) the energy of the ground state (the highest excited state). Both energy scales of the isolated Kitaev chain are plotted in Fig.~\ref{fig:PhotoBands}. The band gap with a green dotted line and the bandwidth with an orange dashed line.  For $\omega_c \ll E_{bw}$, all the energy levels are mixed, and the energy levels from the photonic blocks $n>0$ are localized within the photonic block with $n=0$. On the other hand, for $\omega_c > E_{bw}$ the energy levels corresponding to the photonic block with zero photons $n = 0$ are separated by a gap from the photonic block with $n=1$ and closely follow the many-body spectrum of the isolated Kitaev chain.

\section{Majorana amplitude dependence on cavity frequency}\label{majo_dens_wc}

\begin{figure}[htb]
    \centering
    \includegraphics[width=0.49\textwidth]{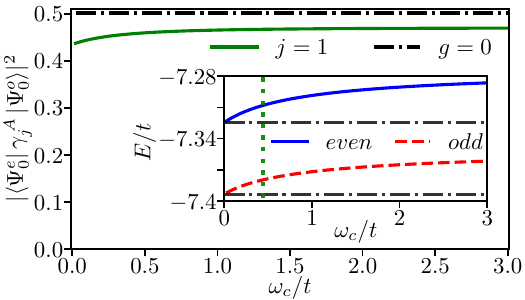}
    \caption{ Amplitude of the Majorana operator $\gamma_j^A=c_j^{\dag}+c_j$ at site $j=1$ between the even $\ket{\Psi^e_0}$ and odd parity $\ket{\Psi^o_0}$ ground states as a function of cavity frequency $\omega_c$. For comparison, the same Majorana amplitude, but for the isolated Kitaev chain ($g=0$) is also plotted in the figure. The inset shows the energy splitting of the ground state with even (blue solid line) and odd parity (red dashed line) as a function of cavity frequency. With dashed-dotted lines we also plot the ground state energies of the isolated Kitaev chain. The parameters of the system are: $\mu/t = 0.75$ $\Delta/t = 0.2$, $g=0.25$, $N=11$, and $N_c = 20$. }
    \label{fig:majo_dens_vs_freq}
\end{figure}

In Fig.~\ref{fig:majo_dens_vs_freq} we define the Majorana operator $\gamma_j^A=c_j^{\dag}+c_j$ to plot the Majorana amplitudes as a function of cavity frequency $\omega_c$ at site $j=1$ between the even $\ket{e}$ and odd parity $\ket{o}$ ground states. An inset is also added to the figure to show the energy dependence of the ground states with $\omega_c$. The ground state energy of the isolated Kitaev chain ($g=0$) is also plotted in the inset with dashed-dotted lines. The gap of the isolated Kitaev chain is also added with a green dotted line. For all the frequencies shown in the figure, the Majorana amplitude in the cavity embedded Kitaev chain is smaller than the Majorana amplitude of the isolated Kitaev chain. We also notice that while the Majorana amplitude is nearly constant for $\omega_c/t \gg \Delta_g$, in the interval $0 < \omega_c < \Delta_g$ the Majorana amplitude slightly increases with increasing frequency. This behavior of the Majorana amplitude with frequency is related to the weak dependence of the ground state energy with frequency, also showing a slight increase with $\omega_c$.

\section{Dependence of the parity switches on $g$}\label{phasediag}

\begin{figure}[htb]
    \centering
    \includegraphics[width=0.49\textwidth]{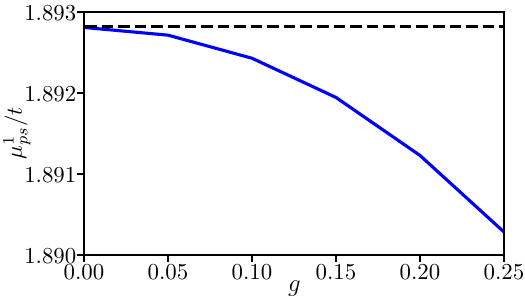}
    \caption{Largest parity switching point as a function of the light-matter coupling strength $g$ for the Kitaev chain coupled to cavity photons (blue solid line) and the isolated Kitaev chain (black dashed line). In the formula for the parity switches of the isolated Kitaev chain, $\mu_{ps} = 2\sqrt{t^2-\Delta^2}\cos\left(p\pi/(N+1)\right)$, the largest parity switching point is obtained when $p=1$. The parameters of the system are: $N=11$, $\mu/t = 1.0$, $\Delta/t = 0.2$, $\omega_c/t = 1.0$, and $N_c = 20$.}
    \label{fig:tran_line}
\end{figure}

In this section, we compute in the Kitaev chain coupled to photons the dependence of the parity switches on $g$ and compare it with the formula for $\mu_{ps}$ of the isolated Kitaev chain. For simplicity we take the largest parity switching point. In the isolated Kitaev chain that corresponds to $p=1$ in the formula $\mu_{ps} = 2\sqrt{t^2-\Delta^2}\cos\left(p\,\pi/(N+1)\right)$. In Fig.~\ref{fig:tran_line} $\mu_{ps}^1$ for the isolated Kitaev chain is plotted with a dashed black line. Notice that the largest parity switching point of the Kitaev chain coupled to cavity photons decreases monotonously with increasing $g$, albeit the decrease is small. This small dependence on $g$ indicates that the formula for parity crossings in the isolated Kitaev chain works as a good approximation for the parity crossings of the Kitaev chain coupled to cavity photons.

\section{Relation between photon number and electronic quantity}\label{photoN_MF}

\begin{figure}[h!]
    \centering
    \includegraphics[width=0.49\textwidth]{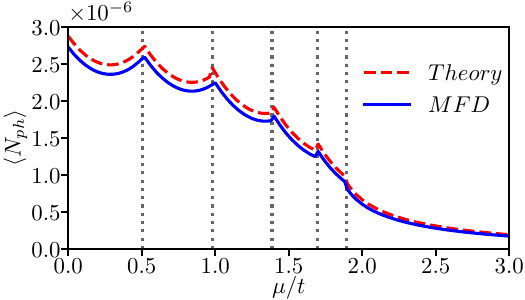}
    \caption{ Photon number obtained with MFD as a function of chemical potential for frequency $\omega_c/t = 5.0$ and light-matter coupling strength $g=0.1$. With a dashed red line we also plot the photon number obtained with the analytical expression~\eqref{eq:NphAnalytical}, derived within the weak coupling regime. Gray vertical lines correspond to the parity switches $\mu_{ps}$. The rest of the parameters of the system are the same as in Fig. \ref{fig:majo_dens_vs_freq}.
}
    \label{fig:PhotonNumberMF}
\end{figure}

In this section, we provide details on the derivation of the expectation value of the photon number based on the MFD between electrons and photons. Expanding the photonic mean-field Hamiltonian $H_{c,ph}^{mf}$ given by Eq.~\eqref{eq:HphMF} to the second order in $g$, we arrive at 

\begin{align}
    H_{c,ph}^{mf} = \omega_c a^\dag a + (a+a^\dag) \langle J_p \rangle- \dfrac{1}{2} (a+a^\dag)^2 \langle J_d\rangle,
    \label{eq:HphMFexpanded}
\end{align}

where 
\begin{align}
	J_p = i \dfrac{g}{\sqrt{N}} \sum_j\Big[- t   c^\dag_{j}c_{j+1} +  \Delta \phi_j c_{j}c_{j+1}- \rm{h.c.}\Big]
	\label{eq:ParamagneticCurrentK}
\end{align}
and 
\begin{align}
	J_d =  \dfrac{g^2}{N} \sum_j\Big[-tc^\dag_jc_{j+1} 
 + \Delta  \phi_j^2  c_{j}c_{j+1}+ \rm{h.c.}\Big],
	\label{eq:DiamagneticCurrentK}
\end{align}
are the paramagnetic and diamagnetic current operators, respectively. We note that $\langle J_p \rangle = 0$ and, therefore, we can bring Eq.~\eqref{eq:HphMFexpanded} to diagonal form $ H_{c,ph}^{mf}  = \omega_c \lambda b^\dag b$ by performing 
the transformation~\cite{Andolina_no-go_theorem}
\begin{align}
a = \cosh(x) b -\sinh(x) b^\dag,
\end{align}
where $\cosh(x) = (\lambda+1)/(2\sqrt{\lambda})$ and $\sinh(x) = (\lambda-1)/(2\sqrt{\lambda})$ with
$\lambda = \sqrt{1-2 \langle J_d\rangle/\omega_c}$.

We assume that $\langle J_d\rangle$ is calculated only over the electronic ground state (and not self-consistently) and we compute the expectation value of the photon number over the ground state of the photonic mean-field Hamiltonian. We arrive at
\begin{align}
    \langle a^\dag a\rangle = \sinh(x)^2\langle bb^\dag \rangle = \dfrac{(\lambda-1)^2}{4\lambda} = \notag\\
    -\frac{\langle J_d\rangle}{2 \omega_c \sqrt{1-\frac{2 \langle J_d\rangle}{\omega_c}}}+\frac{1}{2 \sqrt{1-\frac{2 \langle J_d\rangle}{\omega_c}}}-\frac{1}{2}.
    \label{eq:NphAnalytical}
\end{align}

For large $\omega_c/t\gg 1$ we obtain that the photon number is simplified as
$\langle a^\dag a\rangle \approx \langle J_d\rangle^2/ \left(4 \omega_c^2\right)$ and that it scales as $\propto g^4$.

In Fig.~\ref{fig:PhotonNumberMF} we compare the photon number obtained from Eqs.~\ref{eq:HelMF} and~\ref{eq:HphMF} (MFD) with the analytical expression for the photon number of Eq.~\ref{eq:NphAnalytical}. The peaks in the photon number at the parity switching points are well captured with both methods. Moreover, both methods yield close values for the photon number, with expression~\ref{eq:NphAnalytical} yielding slightly larger values than those of MFD.

\section{Photon number in excited states}\label{PhotoNExc}

%%%%%%%%%%%%%%%%%%%%%%%%%%%%%%%%%%%%%%%%%%%%%
%%%%%%%%%%%%%%%%%%%%%%%%%%%%%%%%%%%%%%%%%%%%%
\begin{figure}{}
	\centering
	\includegraphics[width=0.49\textwidth]{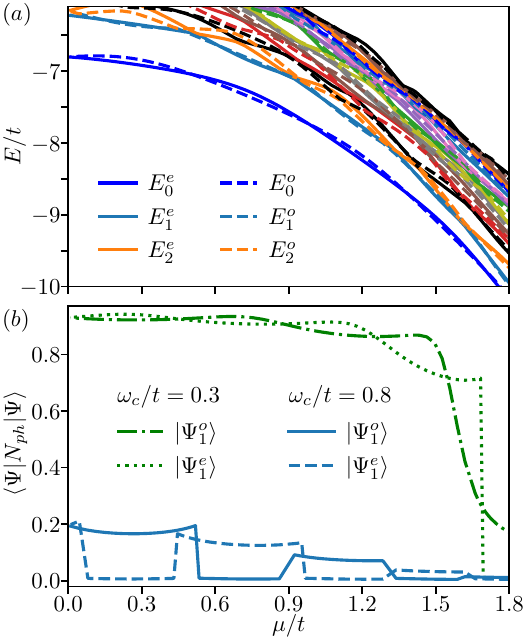}
	\caption{ Many-body energy spectrum of the cavity-embedded Kitaev chain as a function of the chemical potential $\mu$. States with even (odd) parity are plotted with straight (dashed) lines. The cavity frequency considered here is $\omega_c/t = 0.8$. (b) Photon number at excited state as a function of chemical potential for frequencies $\omega_c/t = 0.3,0.8$.  The excited states considered are $\ket{\Psi_1^e}$ and $\ket{\Psi_1^o}$. While, at $\omega_c/t=0.8$ sharp changes in the the photon number are seen, at $\omega_c/t=0.3$ the photon numbers $\bra{\Psi_1^o}N_{ph}\ket{\Psi_1^o}$ and $\bra{\Psi_1^e}N_{ph}\ket{\Psi_1^e}$ are mostly smooth and intertwined. The sharp changes in the photon number of panel (b) stem from higher energies, namely $E_2^e$ and $E_2^o$, approaching closely to the energies $E_1^e$ and $E_1^o$. The parameters of the system are the same as in Fig. \ref{fig:majo_dens_vs_freq}.}
	\label{PhotonNumb:excited}
\end{figure}
%%%%%%%%%%%%%%%%%%%%%%%%%%%%%%%%%%%%%%%%%%%%%
%%%%%%%%%%%%%%%%%%%%%%%%%%%%%%%%%%%%%%%%%%%%%

In this section, by exploiting the advantage that ED has over techniques mainly devoted to the ground state, we compute the photon number at some excited states.  In Fig.~\ref{PhotonNumb:excited}(a) we plot the many-body energy spectrum of Hamiltonian~\eqref{eq:Hc} as a function of chemical potential for $g=0.25$ and $\omega_c/t = 0.8$. Notice the alternation of the parity in the ground state, as already shown in Fig.~\ref{fig:PhotonNumber}, and the  gap separating excited states from the doubly degenerate ground state. In panel (b) we plot the photon number at the excited states $\ket{\Psi_1^e}$ and $\ket{\Psi_1^o}$, having energies $E_1^e$ and $E_1^o$ in panel (a), respectively. At $\omega_c/t = 0.8$ and as a function of $\mu$, we observe an alternant behavior between nonzero and zero values in both $\bra{\Psi_1^o }N_{ph}\ket{\Psi_1^o}$ (solid blue) and $\bra{\Psi_1^e }N_{ph}\ket{\Psi_1^e}$ (dashed blue). Moreover, this alternant behavior is attenuated as $\mu$ increases. There also exists a shift between $\bra{\Psi_1^o }N_{ph}\ket{\Psi_1^o}$ and $\bra{\Psi_1^e }N_{ph}\ket{\Psi_1^e}$ such that when the former is maximum, the latter is minimum. The transitions between the nonzero and zero regions are sharp in both photon numbers. Looking at the energies of the excited states, we notice that the sharp transitions in the photon number at states $\ket{\Psi_1^o}$ and $\ket{\Psi_1^e}$ coincide with the approach of the higher energy states $E_2^o$ and $E_2^e$ to the energies $E_1^o$ and $E_1^e$. The energy spectrum at $\omega_c/t=0.3$ lacks the close approach of the levels $E_2^o$ and $E_2^e$ to the levels $E_1^o$ and $E_1^e$ and therefore does not show sharp jumps in the photon number. 

\section{Cutoff in the number of photonic Fock states}\label{photonCutoff}

\begin{figure}
    \centering
    \includegraphics[width=0.95\linewidth]{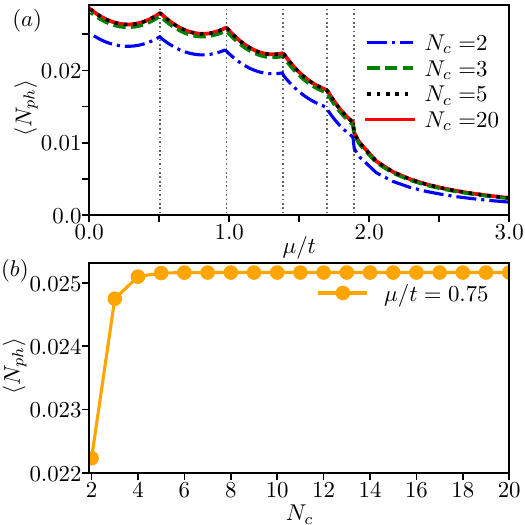}
    \caption{(a) Photon number in the ground state as a function of chemical potential for different cutoffs in the number of photonic Fock states, $N_c$, indicated in the figure.  (b) Photon number in the ground state as a function of $N_c$ for $\mu/t = 0.75$. Photon number remains unchanged for $N_c \geq 5$. Other parameters are fixed as $\Delta/t=0.2$, $\omega_c/t=0.3$, $g=0.25$, and $N=11$.}
    \label{phNumb_coff}
\end{figure}

In this section, we study the dependence of the photon number in the ground state on the photonic cutoff $N_c$. In Fig.~\ref{phNumb_coff} we plot the photon number in the ground state as a function of chemical potential for different $N_c$ [Fig.~\ref{phNumb_coff}~(a)], and as a function of $N_c$ for a given value of chemical potential [Fig.~\ref{phNumb_coff}~(b)]. The photon number $\langle N_{ph}\rangle$ becomes independent on the photonic cutoff for $N_c \geq 5$, justifying the choice for the photonic cutoff in our simulations ($N_c=20$) presented in the rest of the work. 

\section{More about the field quadratures}\label{Quad}

\begin{figure}[htb]
    \centering
    \includegraphics[width=0.49\textwidth]{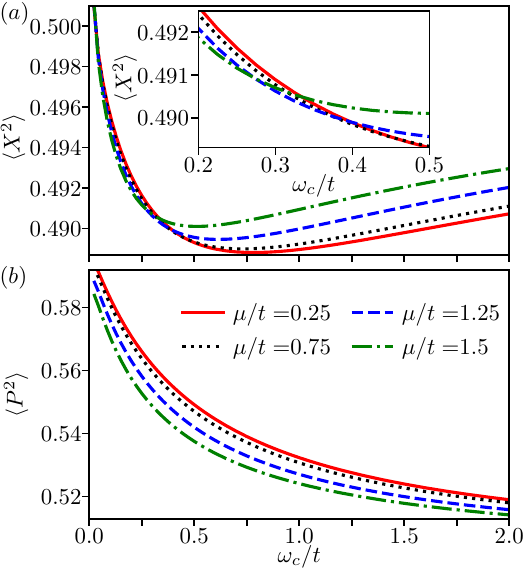}
    \caption{  Photonic quadratures (a) $\langle X^2\rangle$ and (b) $\langle P^2\rangle$ computed at the ground state as a function of cavity frequency $\omega_c/t\,$ for different chemical potentials $\mu/t$, indicated directly in the figure. The quadrature operators are $X=(a+a^{\dag})/\sqrt{2}$ and $P=-i(a-a^{\dag})/\sqrt{2}$. Notice that $\langle P^2\rangle$ decrease with increasing $\omega_c$ independently of $\mu$. On the other hand, $\langle X^2\rangle$ features: 1) a local minimum separating regions with decreasing and increasing order as a function of $\omega_c$, and 2) the value of the minimum and its position in $\omega_c$ changes considerably with $\mu$. Introducing in the decreasing region two threshold frequencies, roughly at $\omega_c = 0.3$ and $0.4$, the inset of panel (a) shows that for $\omega_c < 0.3$ ($\omega_c > 0.4$) $\langle X^2\rangle$ decreases (increases) with increasing $\mu$. The parameters of the system are the same as in Fig. \ref{fig:majo_dens_vs_freq}. 
    }
    \label{Quad:freq}
\end{figure}

\begin{figure}[h!]
    \centering
    \includegraphics[width=0.49\textwidth]{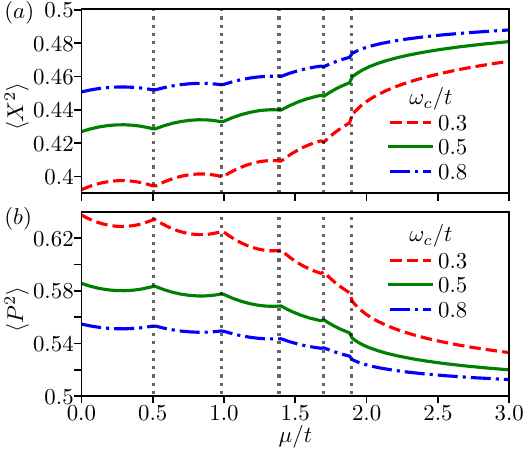}
    \caption{Photonic quadratures (a) $\langle X^2\rangle$ and (b) $\langle P^2\rangle$ obtained by solving the mean field Hamiltonian \eqref{eq:HelMF} and \eqref{eq:HphMF} as a function of chemical potential $\mu/t$ for different cavity frequencies $\omega_c/t$. Gray vertical lines correspond to the parity switches $\mu_{ps}$.  The parameters of the system are the same as in Fig. \ref{fig:majo_dens_vs_freq}. 
    }
    \label{fig:MFQuadratures}
\end{figure}

Introducing the quadrature operators $X=\left(a+a^{\dag}\right)/\sqrt{2}$ and $P=-i\left(a-a^{\dag}\right)/\sqrt{2}$, usually known as the photonic position and momentum operators, we calculate the quadratures $\langle X^2 \rangle$ and $\langle P^2 \rangle$ in the ground state to complement the findings summarized in Fig.~\ref{Quad:ground}. In Fig.~\ref{Quad:freq} the quadratures are plotted as a function of cavity frequency $\omega_c$. The figure shows that the $\langle P^2\rangle$ quadrature monotonically decreases with increasing $\omega_c$, in agreement with the behavior of Fig.~\ref{Quad:ground}~(b) in the main text. On the other hand, the squeezed quadrature $\langle X^2\rangle$ features a minimum that separates decreasing and increasing regions. Notice that increasing $\mu$ not only lifts the minimum but also shifts its position towards smaller frequencies. Importantly, the minimum yields the maximum squeezing in the $\langle X^2\rangle$ quadrature. To describe the dominant characteristics in the squeezed quadrature as a function of $\mu$, let us introduce two threshold frequencies roughly at $\omega_c=0.3$ and $0.4$. The inset in panel (a) shows that for $\omega_c < 0.3$ and ($\omega_c > 0.4$) the $\langle X^2\rangle$ quadrature decreases (increases) with increasing $\mu$. The behavior of $\langle X^2\rangle$ with $\omega_c$ having a minimum is similar to that predicted in photonic superposition states~\cite{squeeze_Wodkiewicz}. 

We can also employ MFD of Appendix~\ref{solvers} to compute the photonic quadratures. They are plotted in Fig.~\ref{fig:MFQuadratures}. Notice that the dips and peaks reported in the $\langle X^2\rangle$ and $\langle P^2\rangle$ quadratures of Fig.~\ref{Quad:ground}, computed from the ED of Hamiltonian~\eqref{eq:Hc}, also appear in the quadratures shown in Fig.~\ref{fig:MFQuadratures}. However, certain differences are worth mentioning. The agreement between ED and MFD for the $\langle P^2\rangle$ quadrature is mostly qualitative given that the latter yields larger values of $\langle P^2\rangle$ than the former. For the squeezed quadrature ($\langle X^2\rangle$), the differences between the two methods are more contrasting. Fig.~\ref{fig:MFQuadratures}(a) suggests that $\langle X^2\rangle$ increases with increasing $\omega_c$ and independently of $\mu$, contrary to Fig.~\ref{Quad:ground}~(a).

\section{Photonic probabilities as a function of cavity frequency}\label{PhProba}

\begin{figure*}[h!]
	\centering
	\includegraphics[width=0.9\textwidth]{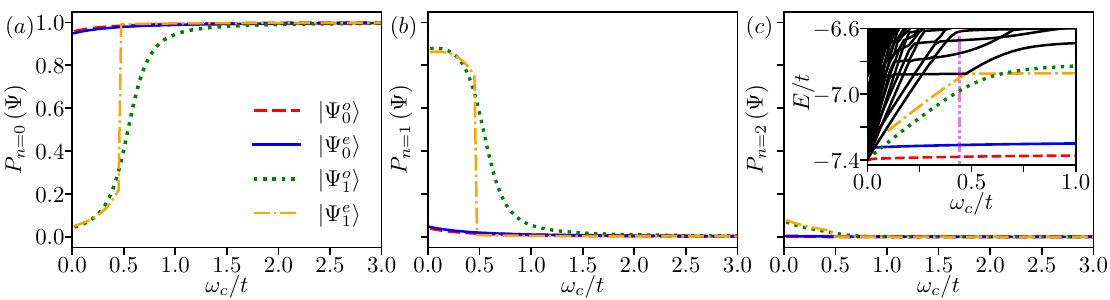}
	\caption{Probability $P_n=|\bra{n} \Psi \rangle|^2$ to find the light-matter state $\ket{\Psi}$ in the photon states $\ket{0}$ (a), $\ket{1}$ (b), and $\ket{2}$ (c), as a function of frequency $\omega_c$. The even and odd parity ground states, corresponding to the states $\ket{\Psi_0^e}$ and $\ket{\Psi_0^o}$, have probabilities $P_{n=0}(\Psi_0^e) \approx  1$ and $P_{n=0}(\Psi_0^o) \approx  1$ that are weakly dependent of frequency. On the other hand, excited states such as $\ket{\Psi_1^e}$ and $\ket{\Psi_1^o}$  show photon probabilities that depend on frequency, especially the probabilities $P_{n=0}(\Psi_1^{e(o)})$ and $P_{n=1}(\Psi_1^{e(o)})$. The inset in panel (c)  shows the many-body energy spectrum as a function of cavity frequency. Moreover, the gap of the isolated Kitaev chain $\Delta_g$ is also plotted in the inset with a pink dash double-dot line. The parameters of the system are the same as in Fig. \ref{fig:majo_dens_vs_freq}.}
    \label{fig:PhProb_supp}
\end{figure*}

In this section, we complement our findings of the photon probability in the ground state and few excited states summarized in Fig.~\ref{fig:PhotonProb}. Defining the photon probability $P_n(\Psi)=|\inpro{n}{\Psi}|^2$, where $\ket{n}$
is a photonic Fock state and $\ket{\Psi}$ a light-matter state, in Fig.~\ref{fig:PhProb_supp} we plot the photon probability as a function of cavity frequency for the ground states $\ket{\Psi_0^e}$ and $\ket{\Psi_0^o}$ and the excited states $\ket{\Psi_1^e}$ and $\ket{\Psi_1^o}$. While the superscripts $e$ and $o$ denote the even and odd parity of the light-matter state, the subscript denotes the label that sorts the energies in increasing order. For the ground states, note that among the photon probabilities shown in panels $(a)$ to $(c)$, the dominant probabilities are $P_{n=0}(\Psi_0^e)$ and $P_{n=0}(\Psi_0^o)$, both approaching one and nearly independent of the cavity frequency. On the other hand, the probabilities in the excited states $\ket{\Psi_1^e}$ and $\ket{\Psi_1^o}$, especially $P_{n=0}(\Psi_1^{e(o)})$ and $P_{n=1}(\Psi_1^{e(o)})$, show a stronger dependence with the cavity frequency. Both photon probabilities feature step-like functions where the probabilities change from nearly zero to nearly one. Moreover, the sharpness of the transitions is different for the two excited states. Interestingly, the transition is located around $\omega_c/t \approx \Delta_g$, the value where the energy spectrum shows anticrossings in the excited states $\ket{\Psi_1^e}$ and $\ket{\Psi_1^o}$ (see the inset of the figure).

\bibliography{bibliographyDraft}

\begin{thebibliography}{82}%
\makeatletter
\providecommand \@ifxundefined [1]{%
 \@ifx{#1\undefined}
}%
\providecommand \@ifnum [1]{%
 \ifnum #1\expandafter \@firstoftwo
 \else \expandafter \@secondoftwo
 \fi
}%
\providecommand \@ifx [1]{%
 \ifx #1\expandafter \@firstoftwo
 \else \expandafter \@secondoftwo
 \fi
}%
\providecommand \natexlab [1]{#1}%
\providecommand \enquote  [1]{``#1''}%
\providecommand \bibnamefont  [1]{#1}%
\providecommand \bibfnamefont [1]{#1}%
\providecommand \citenamefont [1]{#1}%
\providecommand \href@noop [0]{\@secondoftwo}%
\providecommand \href [0]{\begingroup \@sanitize@url \@href}%
\providecommand \@href[1]{\@@startlink{#1}\@@href}%
\providecommand \@@href[1]{\endgroup#1\@@endlink}%
\providecommand \@sanitize@url [0]{\catcode `\\12\catcode `\$12\catcode `\&12\catcode `\#12\catcode `\^12\catcode `\_12\catcode `\%12\relax}%
\providecommand \@@startlink[1]{}%
\providecommand \@@endlink[0]{}%
\providecommand \url  [0]{\begingroup\@sanitize@url \@url }%
\providecommand \@url [1]{\endgroup\@href {#1}{\urlprefix }}%
\providecommand \urlprefix  [0]{URL }%
\providecommand \Eprint [0]{\href }%
\providecommand \doibase [0]{https://doi.org/}%
\providecommand \selectlanguage [0]{\@gobble}%
\providecommand \bibinfo  [0]{\@secondoftwo}%
\providecommand \bibfield  [0]{\@secondoftwo}%
\providecommand \translation [1]{[#1]}%
\providecommand \BibitemOpen [0]{}%
\providecommand \bibitemStop [0]{}%
\providecommand \bibitemNoStop [0]{.\EOS\space}%
\providecommand \EOS [0]{\spacefactor3000\relax}%
\providecommand \BibitemShut  [1]{\csname bibitem#1\endcsname}%
\let\auto@bib@innerbib\@empty
%</preamble>
\bibitem [{\citenamefont {McIver}\ \emph {et~al.}(2020)\citenamefont {McIver}, \citenamefont {Schulte}, \citenamefont {Stein}, \citenamefont {Matsuyama}, \citenamefont {Jotzu}, \citenamefont {Meier},\ and\ \citenamefont {Cavalleri}}]{McIver_exp_AHE}%
  \BibitemOpen
  \bibfield  {author} {\bibinfo {author} {\bibfnamefont {J.~W.}\ \bibnamefont {McIver}}, \bibinfo {author} {\bibfnamefont {B.}~\bibnamefont {Schulte}}, \bibinfo {author} {\bibfnamefont {F.-U.}\ \bibnamefont {Stein}}, \bibinfo {author} {\bibfnamefont {T.}~\bibnamefont {Matsuyama}}, \bibinfo {author} {\bibfnamefont {G.}~\bibnamefont {Jotzu}}, \bibinfo {author} {\bibfnamefont {G.}~\bibnamefont {Meier}},\ and\ \bibinfo {author} {\bibfnamefont {A.}~\bibnamefont {Cavalleri}},\ }\bibfield  {title} {\bibinfo {title} {Light-induced anomalous {H}all effect in graphene},\ }\href {https://doi.org/10.1038/s41567-019-0698-y} {\bibfield  {journal} {\bibinfo  {journal} {Nature Physics}\ }\textbf {\bibinfo {volume} {16}},\ \bibinfo {pages} {38} (\bibinfo {year} {2020})}\BibitemShut {NoStop}%
\bibitem [{\citenamefont {Wang}\ \emph {et~al.}(2013)\citenamefont {Wang}, \citenamefont {Steinberg}, \citenamefont {Jarillo-Herrero},\ and\ \citenamefont {Gedik}}]{Wang_Floquet_BiSe}%
  \BibitemOpen
  \bibfield  {author} {\bibinfo {author} {\bibfnamefont {Y.~H.}\ \bibnamefont {Wang}}, \bibinfo {author} {\bibfnamefont {H.}~\bibnamefont {Steinberg}}, \bibinfo {author} {\bibfnamefont {P.}~\bibnamefont {Jarillo-Herrero}},\ and\ \bibinfo {author} {\bibfnamefont {N.}~\bibnamefont {Gedik}},\ }\bibfield  {title} {\bibinfo {title} {Observation of {F}loquet-{B}loch states on the surface of a topological insulator},\ }\href {https://doi.org/10.1126/science.1239834} {\bibfield  {journal} {\bibinfo  {journal} {Science}\ }\textbf {\bibinfo {volume} {342}},\ \bibinfo {pages} {453} (\bibinfo {year} {2013})},\ \Eprint {https://arxiv.org/abs/https://www.science.org/doi/pdf/10.1126/science.1239834} {https://www.science.org/doi/pdf/10.1126/science.1239834} \BibitemShut {NoStop}%
\bibitem [{\citenamefont {Mitrano}\ \emph {et~al.}(2016)\citenamefont {Mitrano}, \citenamefont {Cantaluppi}, \citenamefont {Nicoletti}, \citenamefont {Kaiser}, \citenamefont {Perucchi}, \citenamefont {Lupi}, \citenamefont {Di~Pietro}, \citenamefont {Pontiroli}, \citenamefont {Ricc{\`o}}, \citenamefont {Clark}, \citenamefont {Jaksch},\ and\ \citenamefont {Cavalleri}}]{Mitrano_LightInd_SC}%
  \BibitemOpen
  \bibfield  {author} {\bibinfo {author} {\bibfnamefont {M.}~\bibnamefont {Mitrano}}, \bibinfo {author} {\bibfnamefont {A.}~\bibnamefont {Cantaluppi}}, \bibinfo {author} {\bibfnamefont {D.}~\bibnamefont {Nicoletti}}, \bibinfo {author} {\bibfnamefont {S.}~\bibnamefont {Kaiser}}, \bibinfo {author} {\bibfnamefont {A.}~\bibnamefont {Perucchi}}, \bibinfo {author} {\bibfnamefont {S.}~\bibnamefont {Lupi}}, \bibinfo {author} {\bibfnamefont {P.}~\bibnamefont {Di~Pietro}}, \bibinfo {author} {\bibfnamefont {D.}~\bibnamefont {Pontiroli}}, \bibinfo {author} {\bibfnamefont {M.}~\bibnamefont {Ricc{\`o}}}, \bibinfo {author} {\bibfnamefont {S.~R.}\ \bibnamefont {Clark}}, \bibinfo {author} {\bibfnamefont {D.}~\bibnamefont {Jaksch}},\ and\ \bibinfo {author} {\bibfnamefont {A.}~\bibnamefont {Cavalleri}},\ }\bibfield  {title} {\bibinfo {title} {Possible light-induced superconductivity in k3c60 at high temperature},\ }\href {https://doi.org/10.1038/nature16522} {\bibfield  {journal} {\bibinfo  {journal} {Nature}\ }\textbf
  {\bibinfo {volume} {530}},\ \bibinfo {pages} {461} (\bibinfo {year} {2016})}\BibitemShut {NoStop}%
\bibitem [{\citenamefont {Budden}\ \emph {et~al.}(2021)\citenamefont {Budden}, \citenamefont {Gebert}, \citenamefont {Buzzi}, \citenamefont {Jotzu}, \citenamefont {Wang}, \citenamefont {Matsuyama}, \citenamefont {Meier}, \citenamefont {Laplace}, \citenamefont {Pontiroli}, \citenamefont {Ricc{\`o}}, \citenamefont {Schlawin}, \citenamefont {Jaksch},\ and\ \citenamefont {Cavalleri}}]{Budden_LightInd_SC}%
  \BibitemOpen
  \bibfield  {author} {\bibinfo {author} {\bibfnamefont {M.}~\bibnamefont {Budden}}, \bibinfo {author} {\bibfnamefont {T.}~\bibnamefont {Gebert}}, \bibinfo {author} {\bibfnamefont {M.}~\bibnamefont {Buzzi}}, \bibinfo {author} {\bibfnamefont {G.}~\bibnamefont {Jotzu}}, \bibinfo {author} {\bibfnamefont {E.}~\bibnamefont {Wang}}, \bibinfo {author} {\bibfnamefont {T.}~\bibnamefont {Matsuyama}}, \bibinfo {author} {\bibfnamefont {G.}~\bibnamefont {Meier}}, \bibinfo {author} {\bibfnamefont {Y.}~\bibnamefont {Laplace}}, \bibinfo {author} {\bibfnamefont {D.}~\bibnamefont {Pontiroli}}, \bibinfo {author} {\bibfnamefont {M.}~\bibnamefont {Ricc{\`o}}}, \bibinfo {author} {\bibfnamefont {F.}~\bibnamefont {Schlawin}}, \bibinfo {author} {\bibfnamefont {D.}~\bibnamefont {Jaksch}},\ and\ \bibinfo {author} {\bibfnamefont {A.}~\bibnamefont {Cavalleri}},\ }\bibfield  {title} {\bibinfo {title} {Evidence for metastable photo-induced superconductivity in k3c60},\ }\href {https://doi.org/10.1038/s41567-020-01148-1} {\bibfield
  {journal} {\bibinfo  {journal} {Nature Physics}\ }\textbf {\bibinfo {volume} {17}},\ \bibinfo {pages} {611} (\bibinfo {year} {2021})}\BibitemShut {NoStop}%
\bibitem [{\citenamefont {Schlawin}\ \emph {et~al.}(2022)\citenamefont {Schlawin}, \citenamefont {Kennes},\ and\ \citenamefont {Sentef}}]{CavRevw_Schlawin}%
  \BibitemOpen
  \bibfield  {author} {\bibinfo {author} {\bibfnamefont {F.}~\bibnamefont {Schlawin}}, \bibinfo {author} {\bibfnamefont {D.~M.}\ \bibnamefont {Kennes}},\ and\ \bibinfo {author} {\bibfnamefont {M.~A.}\ \bibnamefont {Sentef}},\ }\bibfield  {title} {\bibinfo {title} {Cavity quantum materials},\ }\href {https://doi.org/10.1063/5.0083825} {\bibfield  {journal} {\bibinfo  {journal} {Applied Physics Reviews}\ }\textbf {\bibinfo {volume} {9}},\ \bibinfo {pages} {011312} (\bibinfo {year} {2022})},\ \Eprint {https://arxiv.org/abs/https://pubs.aip.org/aip/apr/article-pdf/doi/10.1063/5.0083825/19819541/011312\_1\_online.pdf} {https://pubs.aip.org/aip/apr/article-pdf/doi/10.1063/5.0083825/19819541/011312\_1\_online.pdf} \BibitemShut {NoStop}%
\bibitem [{\citenamefont {Appugliese}\ \emph {et~al.}(2022)\citenamefont {Appugliese}, \citenamefont {Enkner}, \citenamefont {Paravicini-Bagliani}, \citenamefont {Beck}, \citenamefont {Reichl}, \citenamefont {Wegscheider}, \citenamefont {Scalari}, \citenamefont {Ciuti},\ and\ \citenamefont {Faist}}]{appugliese2022breakdown}%
  \BibitemOpen
  \bibfield  {author} {\bibinfo {author} {\bibfnamefont {F.}~\bibnamefont {Appugliese}}, \bibinfo {author} {\bibfnamefont {J.}~\bibnamefont {Enkner}}, \bibinfo {author} {\bibfnamefont {G.~L.}\ \bibnamefont {Paravicini-Bagliani}}, \bibinfo {author} {\bibfnamefont {M.}~\bibnamefont {Beck}}, \bibinfo {author} {\bibfnamefont {C.}~\bibnamefont {Reichl}}, \bibinfo {author} {\bibfnamefont {W.}~\bibnamefont {Wegscheider}}, \bibinfo {author} {\bibfnamefont {G.}~\bibnamefont {Scalari}}, \bibinfo {author} {\bibfnamefont {C.}~\bibnamefont {Ciuti}},\ and\ \bibinfo {author} {\bibfnamefont {J.}~\bibnamefont {Faist}},\ }\bibfield  {title} {\bibinfo {title} {Breakdown of topological protection by cavity vacuum fields in the integer quantum {H}all effect},\ }\href {https://doi.org/10.1126/science.abl5818} {\bibfield  {journal} {\bibinfo  {journal} {Science}\ }\textbf {\bibinfo {volume} {375}},\ \bibinfo {pages} {1030} (\bibinfo {year} {2022})}\BibitemShut {NoStop}%
\bibitem [{\citenamefont {Jarc}\ \emph {et~al.}(2023)\citenamefont {Jarc}, \citenamefont {Mathengattil}, \citenamefont {Montanaro}, \citenamefont {Giusti}, \citenamefont {Rigoni}, \citenamefont {Sergo}, \citenamefont {Fassioli}, \citenamefont {Winnerl}, \citenamefont {Dal~Zilio}, \citenamefont {Mihailovic}, \citenamefont {Prelov{\v{s}}ek}, \citenamefont {Eckstein},\ and\ \citenamefont {Fausti}}]{Jarc_exp_CDW}%
  \BibitemOpen
  \bibfield  {author} {\bibinfo {author} {\bibfnamefont {G.}~\bibnamefont {Jarc}}, \bibinfo {author} {\bibfnamefont {S.~Y.}\ \bibnamefont {Mathengattil}}, \bibinfo {author} {\bibfnamefont {A.}~\bibnamefont {Montanaro}}, \bibinfo {author} {\bibfnamefont {F.}~\bibnamefont {Giusti}}, \bibinfo {author} {\bibfnamefont {E.~M.}\ \bibnamefont {Rigoni}}, \bibinfo {author} {\bibfnamefont {R.}~\bibnamefont {Sergo}}, \bibinfo {author} {\bibfnamefont {F.}~\bibnamefont {Fassioli}}, \bibinfo {author} {\bibfnamefont {S.}~\bibnamefont {Winnerl}}, \bibinfo {author} {\bibfnamefont {S.}~\bibnamefont {Dal~Zilio}}, \bibinfo {author} {\bibfnamefont {D.}~\bibnamefont {Mihailovic}}, \bibinfo {author} {\bibfnamefont {P.}~\bibnamefont {Prelov{\v{s}}ek}}, \bibinfo {author} {\bibfnamefont {M.}~\bibnamefont {Eckstein}},\ and\ \bibinfo {author} {\bibfnamefont {D.}~\bibnamefont {Fausti}},\ }\bibfield  {title} {\bibinfo {title} {Cavity-mediated thermal control of metal-to-insulator transition in 1t-tas2},\ }\href
  {https://doi.org/10.1038/s41586-023-06596-2} {\bibfield  {journal} {\bibinfo  {journal} {Nature}\ }\textbf {\bibinfo {volume} {622}},\ \bibinfo {pages} {487} (\bibinfo {year} {2023})}\BibitemShut {NoStop}%
\bibitem [{\citenamefont {Hasan}\ and\ \citenamefont {Kane}(2010)}]{TopoMatter_Rev_Hasan}%
  \BibitemOpen
  \bibfield  {author} {\bibinfo {author} {\bibfnamefont {M.~Z.}\ \bibnamefont {Hasan}}\ and\ \bibinfo {author} {\bibfnamefont {C.~L.}\ \bibnamefont {Kane}},\ }\bibfield  {title} {\bibinfo {title} {Colloquium: Topological insulators},\ }\href {https://doi.org/10.1103/RevModPhys.82.3045} {\bibfield  {journal} {\bibinfo  {journal} {Rev. Mod. Phys.}\ }\textbf {\bibinfo {volume} {82}},\ \bibinfo {pages} {3045} (\bibinfo {year} {2010})}\BibitemShut {NoStop}%
\bibitem [{\citenamefont {Qi}\ and\ \citenamefont {Zhang}(2011)}]{TopoMatter_Rev_Qi}%
  \BibitemOpen
  \bibfield  {author} {\bibinfo {author} {\bibfnamefont {X.-L.}\ \bibnamefont {Qi}}\ and\ \bibinfo {author} {\bibfnamefont {S.-C.}\ \bibnamefont {Zhang}},\ }\bibfield  {title} {\bibinfo {title} {Topological insulators and superconductors},\ }\href {https://doi.org/10.1103/RevModPhys.83.1057} {\bibfield  {journal} {\bibinfo  {journal} {Rev. Mod. Phys.}\ }\textbf {\bibinfo {volume} {83}},\ \bibinfo {pages} {1057} (\bibinfo {year} {2011})}\BibitemShut {NoStop}%
\bibitem [{\citenamefont {Sato}\ and\ \citenamefont {Ando}(2017)}]{TopoSup_Rev_Sato}%
  \BibitemOpen
  \bibfield  {author} {\bibinfo {author} {\bibfnamefont {M.}~\bibnamefont {Sato}}\ and\ \bibinfo {author} {\bibfnamefont {Y.}~\bibnamefont {Ando}},\ }\bibfield  {title} {\bibinfo {title} {Topological superconductors: a review},\ }\href {https://doi.org/10.1088/1361-6633/aa6ac7} {\bibfield  {journal} {\bibinfo  {journal} {Reports on Progress in Physics}\ }\textbf {\bibinfo {volume} {80}},\ \bibinfo {pages} {076501} (\bibinfo {year} {2017})}\BibitemShut {NoStop}%
\bibitem [{\citenamefont {Kitaev}(2001)}]{kitaev2001unpaired}%
  \BibitemOpen
  \bibfield  {author} {\bibinfo {author} {\bibfnamefont {A.~Y.}\ \bibnamefont {Kitaev}},\ }\bibfield  {title} {\bibinfo {title} {Unpaired {M}ajorana fermions in quantum wires},\ }\href {https://doi.org/10.1070/1063-7869/44/10S/S29} {\bibfield  {journal} {\bibinfo  {journal} {Physics-Uspekhi}\ }\textbf {\bibinfo {volume} {44}},\ \bibinfo {pages} {131} (\bibinfo {year} {2001})}\BibitemShut {NoStop}%
\bibitem [{\citenamefont {Su}\ \emph {et~al.}(1979)\citenamefont {Su}, \citenamefont {Schrieffer},\ and\ \citenamefont {Heeger}}]{SSH_paper}%
  \BibitemOpen
  \bibfield  {author} {\bibinfo {author} {\bibfnamefont {W.~P.}\ \bibnamefont {Su}}, \bibinfo {author} {\bibfnamefont {J.~R.}\ \bibnamefont {Schrieffer}},\ and\ \bibinfo {author} {\bibfnamefont {A.~J.}\ \bibnamefont {Heeger}},\ }\bibfield  {title} {\bibinfo {title} {Solitons in polyacetylene},\ }\href {https://doi.org/10.1103/PhysRevLett.42.1698} {\bibfield  {journal} {\bibinfo  {journal} {Phys. Rev. Lett.}\ }\textbf {\bibinfo {volume} {42}},\ \bibinfo {pages} {1698} (\bibinfo {year} {1979})}\BibitemShut {NoStop}%
\bibitem [{\citenamefont {Lutchyn}\ \emph {et~al.}(2010)\citenamefont {Lutchyn}, \citenamefont {Sau},\ and\ \citenamefont {Das~Sarma}}]{Majo_wires_Lutchyn}%
  \BibitemOpen
  \bibfield  {author} {\bibinfo {author} {\bibfnamefont {R.~M.}\ \bibnamefont {Lutchyn}}, \bibinfo {author} {\bibfnamefont {J.~D.}\ \bibnamefont {Sau}},\ and\ \bibinfo {author} {\bibfnamefont {S.}~\bibnamefont {Das~Sarma}},\ }\bibfield  {title} {\bibinfo {title} {Majorana fermions and a topological phase transition in semiconductor-superconductor heterostructures},\ }\href {https://doi.org/10.1103/PhysRevLett.105.077001} {\bibfield  {journal} {\bibinfo  {journal} {Phys. Rev. Lett.}\ }\textbf {\bibinfo {volume} {105}},\ \bibinfo {pages} {077001} (\bibinfo {year} {2010})}\BibitemShut {NoStop}%
\bibitem [{\citenamefont {Oreg}\ \emph {et~al.}(2010)\citenamefont {Oreg}, \citenamefont {Refael},\ and\ \citenamefont {von Oppen}}]{Majo_wires_Oreg}%
  \BibitemOpen
  \bibfield  {author} {\bibinfo {author} {\bibfnamefont {Y.}~\bibnamefont {Oreg}}, \bibinfo {author} {\bibfnamefont {G.}~\bibnamefont {Refael}},\ and\ \bibinfo {author} {\bibfnamefont {F.}~\bibnamefont {von Oppen}},\ }\bibfield  {title} {\bibinfo {title} {Helical liquids and {M}ajorana bound states in quantum wires},\ }\href {https://doi.org/10.1103/PhysRevLett.105.177002} {\bibfield  {journal} {\bibinfo  {journal} {Phys. Rev. Lett.}\ }\textbf {\bibinfo {volume} {105}},\ \bibinfo {pages} {177002} (\bibinfo {year} {2010})}\BibitemShut {NoStop}%
\bibitem [{\citenamefont {Sticlet}\ \emph {et~al.}(2014)\citenamefont {Sticlet}, \citenamefont {Seabra}, \citenamefont {Pollmann},\ and\ \citenamefont {Cayssol}}]{CreutzMajorana}%
  \BibitemOpen
  \bibfield  {author} {\bibinfo {author} {\bibfnamefont {D.}~\bibnamefont {Sticlet}}, \bibinfo {author} {\bibfnamefont {L.}~\bibnamefont {Seabra}}, \bibinfo {author} {\bibfnamefont {F.}~\bibnamefont {Pollmann}},\ and\ \bibinfo {author} {\bibfnamefont {J.}~\bibnamefont {Cayssol}},\ }\bibfield  {title} {\bibinfo {title} {From fractionally charged solitons to {M}ajorana bound states in a one-dimensional interacting model},\ }\href {https://doi.org/10.1103/PhysRevB.89.115430} {\bibfield  {journal} {\bibinfo  {journal} {Phys. Rev. B}\ }\textbf {\bibinfo {volume} {89}},\ \bibinfo {pages} {115430} (\bibinfo {year} {2014})}\BibitemShut {NoStop}%
\bibitem [{\citenamefont {Trif}\ and\ \citenamefont {Tserkovnyak}(2012)}]{trif2012resonantly}%
  \BibitemOpen
  \bibfield  {author} {\bibinfo {author} {\bibfnamefont {M.}~\bibnamefont {Trif}}\ and\ \bibinfo {author} {\bibfnamefont {Y.}~\bibnamefont {Tserkovnyak}},\ }\bibfield  {title} {\bibinfo {title} {Resonantly tunable {M}ajorana polariton in a microwave cavity},\ }\href {https://doi.org/10.1103/PhysRevLett.109.257002} {\bibfield  {journal} {\bibinfo  {journal} {Phys. Rev. Lett.}\ }\textbf {\bibinfo {volume} {109}},\ \bibinfo {pages} {257002} (\bibinfo {year} {2012})}\BibitemShut {NoStop}%
\bibitem [{\citenamefont {Bacciconi}\ \emph {et~al.}(2024)\citenamefont {Bacciconi}, \citenamefont {Andolina},\ and\ \citenamefont {Mora}}]{bacciconi2023topological}%
  \BibitemOpen
  \bibfield  {author} {\bibinfo {author} {\bibfnamefont {Z.}~\bibnamefont {Bacciconi}}, \bibinfo {author} {\bibfnamefont {G.~M.}\ \bibnamefont {Andolina}},\ and\ \bibinfo {author} {\bibfnamefont {C.}~\bibnamefont {Mora}},\ }\bibfield  {title} {\bibinfo {title} {Topological protection of {M}ajorana polaritons in a cavity},\ }\href {https://doi.org/10.1103/PhysRevB.109.165434} {\bibfield  {journal} {\bibinfo  {journal} {Phys. Rev. B}\ }\textbf {\bibinfo {volume} {109}},\ \bibinfo {pages} {165434} (\bibinfo {year} {2024})}\BibitemShut {NoStop}%
\bibitem [{\citenamefont {Dmytruk}\ and\ \citenamefont {Schir\`o}(2024)}]{dmytruk2024hybrid}%
  \BibitemOpen
  \bibfield  {author} {\bibinfo {author} {\bibfnamefont {O.}~\bibnamefont {Dmytruk}}\ and\ \bibinfo {author} {\bibfnamefont {M.}~\bibnamefont {Schir\`o}},\ }\bibfield  {title} {\bibinfo {title} {Hybrid light-matter states in topological superconductors coupled to cavity photons},\ }\href {https://doi.org/10.1103/PhysRevB.110.075416} {\bibfield  {journal} {\bibinfo  {journal} {Phys. Rev. B}\ }\textbf {\bibinfo {volume} {110}},\ \bibinfo {pages} {075416} (\bibinfo {year} {2024})}\BibitemShut {NoStop}%
\bibitem [{\citenamefont {Ciuti}(2021)}]{ciuti2021cavity}%
  \BibitemOpen
  \bibfield  {author} {\bibinfo {author} {\bibfnamefont {C.}~\bibnamefont {Ciuti}},\ }\bibfield  {title} {\bibinfo {title} {Cavity-mediated electron hopping in disordered quantum {H}all systems},\ }\href {https://doi.org/10.1103/PhysRevB.104.155307} {\bibfield  {journal} {\bibinfo  {journal} {Phys. Rev. B}\ }\textbf {\bibinfo {volume} {104}},\ \bibinfo {pages} {155307} (\bibinfo {year} {2021})}\BibitemShut {NoStop}%
\bibitem [{\citenamefont {Dmytruk}\ and\ \citenamefont {Schir{\`o}}(2022)}]{dmytruk2022controlling}%
  \BibitemOpen
  \bibfield  {author} {\bibinfo {author} {\bibfnamefont {O.}~\bibnamefont {Dmytruk}}\ and\ \bibinfo {author} {\bibfnamefont {M.}~\bibnamefont {Schir{\`o}}},\ }\bibfield  {title} {\bibinfo {title} {Controlling topological phases of matter with quantum light},\ }\href {https://doi.org/10.1038/s42005-022-01049-0} {\bibfield  {journal} {\bibinfo  {journal} {Communications Physics}\ }\textbf {\bibinfo {volume} {5}},\ \bibinfo {pages} {271} (\bibinfo {year} {2022})}\BibitemShut {NoStop}%
\bibitem [{\citenamefont {P{\'{e}}rez-Gonz{\'{a}}lez}\ \emph {et~al.}(2025)\citenamefont {P{\'{e}}rez-Gonz{\'{a}}lez}, \citenamefont {Platero},\ and\ \citenamefont {Gomez-Le{\'{o}}n}}]{perez2023light}%
  \BibitemOpen
  \bibfield  {author} {\bibinfo {author} {\bibfnamefont {B.}~\bibnamefont {P{\'{e}}rez-Gonz{\'{a}}lez}}, \bibinfo {author} {\bibfnamefont {G.}~\bibnamefont {Platero}},\ and\ \bibinfo {author} {\bibfnamefont {{\'{A}}.}~\bibnamefont {Gomez-Le{\'{o}}n}},\ }\bibfield  {title} {\bibinfo {title} {Light-matter correlations in {Q}uantum {F}loquet engineering of cavity quantum materials},\ }\href {https://doi.org/10.22331/q-2025-02-17-1633} {\bibfield  {journal} {\bibinfo  {journal} {{Quantum}}\ }\textbf {\bibinfo {volume} {9}},\ \bibinfo {pages} {1633} (\bibinfo {year} {2025})}\BibitemShut {NoStop}%
\bibitem [{\citenamefont {Shaffer}\ \emph {et~al.}(2024)\citenamefont {Shaffer}, \citenamefont {Claassen}, \citenamefont {Srivastava},\ and\ \citenamefont {Santos}}]{shaffer2023entanglement}%
  \BibitemOpen
  \bibfield  {author} {\bibinfo {author} {\bibfnamefont {D.}~\bibnamefont {Shaffer}}, \bibinfo {author} {\bibfnamefont {M.}~\bibnamefont {Claassen}}, \bibinfo {author} {\bibfnamefont {A.}~\bibnamefont {Srivastava}},\ and\ \bibinfo {author} {\bibfnamefont {L.~H.}\ \bibnamefont {Santos}},\ }\bibfield  {title} {\bibinfo {title} {Entanglement and topology in {S}u-{S}chrieffer-{H}eeger cavity quantum electrodynamics},\ }\href {https://doi.org/10.1103/PhysRevB.109.155160} {\bibfield  {journal} {\bibinfo  {journal} {Phys. Rev. B}\ }\textbf {\bibinfo {volume} {109}},\ \bibinfo {pages} {155160} (\bibinfo {year} {2024})}\BibitemShut {NoStop}%
\bibitem [{\citenamefont {Pérez-González}\ \emph {et~al.}(2023)\citenamefont {Pérez-González}, \citenamefont {Platero},\ and\ \citenamefont {Álvaro Gómez-León}}]{perez2023many}%
  \BibitemOpen
  \bibfield  {author} {\bibinfo {author} {\bibfnamefont {B.}~\bibnamefont {Pérez-González}}, \bibinfo {author} {\bibfnamefont {G.}~\bibnamefont {Platero}},\ and\ \bibinfo {author} {\bibnamefont {Álvaro Gómez-León}},\ }\href {https://arxiv.org/abs/2312.10141} {\bibinfo {title} {Many-body origin of anomalous {F}loquet phases in cavity-{QED} materials}} (\bibinfo {year} {2023}),\ \Eprint {https://arxiv.org/abs/2312.10141} {arXiv:2312.10141 [quant-ph]} \BibitemShut {NoStop}%
\bibitem [{\citenamefont {Nguyen}\ \emph {et~al.}(2023)\citenamefont {Nguyen}, \citenamefont {Arwas}, \citenamefont {Lin}, \citenamefont {Yao},\ and\ \citenamefont {Ciuti}}]{nguyen2023electron}%
  \BibitemOpen
  \bibfield  {author} {\bibinfo {author} {\bibfnamefont {D.-P.}\ \bibnamefont {Nguyen}}, \bibinfo {author} {\bibfnamefont {G.}~\bibnamefont {Arwas}}, \bibinfo {author} {\bibfnamefont {Z.}~\bibnamefont {Lin}}, \bibinfo {author} {\bibfnamefont {W.}~\bibnamefont {Yao}},\ and\ \bibinfo {author} {\bibfnamefont {C.}~\bibnamefont {Ciuti}},\ }\bibfield  {title} {\bibinfo {title} {Electron-photon chern number in cavity-embedded {2D} moir\'e materials},\ }\href {https://doi.org/10.1103/PhysRevLett.131.176602} {\bibfield  {journal} {\bibinfo  {journal} {Phys. Rev. Lett.}\ }\textbf {\bibinfo {volume} {131}},\ \bibinfo {pages} {176602} (\bibinfo {year} {2023})}\BibitemShut {NoStop}%
\bibitem [{\citenamefont {Nguyen}\ \emph {et~al.}(2024)\citenamefont {Nguyen}, \citenamefont {Arwas},\ and\ \citenamefont {Ciuti}}]{nguyen2024electron}%
  \BibitemOpen
  \bibfield  {author} {\bibinfo {author} {\bibfnamefont {D.-P.}\ \bibnamefont {Nguyen}}, \bibinfo {author} {\bibfnamefont {G.}~\bibnamefont {Arwas}},\ and\ \bibinfo {author} {\bibfnamefont {C.}~\bibnamefont {Ciuti}},\ }\bibfield  {title} {\bibinfo {title} {Electron conductance and many-body marker of a cavity-embedded topological one-dimensional chain},\ }\href {https://doi.org/10.1103/PhysRevB.110.195416} {\bibfield  {journal} {\bibinfo  {journal} {Phys. Rev. B}\ }\textbf {\bibinfo {volume} {110}},\ \bibinfo {pages} {195416} (\bibinfo {year} {2024})}\BibitemShut {NoStop}%
\bibitem [{\citenamefont {Shaw}\ \emph {et~al.}(2024)\citenamefont {Shaw}, \citenamefont {Hsu},\ and\ \citenamefont {You}}]{shaw2024theoreticalstudycavitymodulatedtopological}%
  \BibitemOpen
  \bibfield  {author} {\bibinfo {author} {\bibfnamefont {Y.-C.}\ \bibnamefont {Shaw}}, \bibinfo {author} {\bibfnamefont {H.-C.}\ \bibnamefont {Hsu}},\ and\ \bibinfo {author} {\bibfnamefont {J.~S.}\ \bibnamefont {You}},\ }\href {https://arxiv.org/abs/2412.19508} {\bibinfo {title} {A theoretical study of cavity-modulated topological {A}nderson insulators}} (\bibinfo {year} {2024}),\ \Eprint {https://arxiv.org/abs/2412.19508} {arXiv:2412.19508 [cond-mat.mtrl-sci]} \BibitemShut {NoStop}%
\bibitem [{\citenamefont {Yavilberg}\ \emph {et~al.}(2015)\citenamefont {Yavilberg}, \citenamefont {Ginossar},\ and\ \citenamefont {Grosfeld}}]{Z2Qubit_Yavilberg}%
  \BibitemOpen
  \bibfield  {author} {\bibinfo {author} {\bibfnamefont {K.}~\bibnamefont {Yavilberg}}, \bibinfo {author} {\bibfnamefont {E.}~\bibnamefont {Ginossar}},\ and\ \bibinfo {author} {\bibfnamefont {E.}~\bibnamefont {Grosfeld}},\ }\bibfield  {title} {\bibinfo {title} {Fermion parity measurement and control in {M}ajorana circuit quantum electrodynamics},\ }\href {https://doi.org/10.1103/PhysRevB.92.075143} {\bibfield  {journal} {\bibinfo  {journal} {Phys. Rev. B}\ }\textbf {\bibinfo {volume} {92}},\ \bibinfo {pages} {075143} (\bibinfo {year} {2015})}\BibitemShut {NoStop}%
\bibitem [{\citenamefont {Trif}\ and\ \citenamefont {Simon}(2019)}]{trif2019braiding}%
  \BibitemOpen
  \bibfield  {author} {\bibinfo {author} {\bibfnamefont {M.}~\bibnamefont {Trif}}\ and\ \bibinfo {author} {\bibfnamefont {P.}~\bibnamefont {Simon}},\ }\bibfield  {title} {\bibinfo {title} {Braiding of {M}ajorana fermions in a cavity},\ }\href {https://doi.org/10.1103/PhysRevLett.122.236803} {\bibfield  {journal} {\bibinfo  {journal} {Phys. Rev. Lett.}\ }\textbf {\bibinfo {volume} {122}},\ \bibinfo {pages} {236803} (\bibinfo {year} {2019})}\BibitemShut {NoStop}%
\bibitem [{\citenamefont {Contamin}\ \emph {et~al.}(2021)\citenamefont {Contamin}, \citenamefont {Delbecq}, \citenamefont {Dou{\c c}ot}, \citenamefont {Cottet},\ and\ \citenamefont {Kontos}}]{contamin2021topological}%
  \BibitemOpen
  \bibfield  {author} {\bibinfo {author} {\bibfnamefont {L.~C.}\ \bibnamefont {Contamin}}, \bibinfo {author} {\bibfnamefont {M.~R.}\ \bibnamefont {Delbecq}}, \bibinfo {author} {\bibfnamefont {B.}~\bibnamefont {Dou{\c c}ot}}, \bibinfo {author} {\bibfnamefont {A.}~\bibnamefont {Cottet}},\ and\ \bibinfo {author} {\bibfnamefont {T.}~\bibnamefont {Kontos}},\ }\bibfield  {title} {\bibinfo {title} {Hybrid light-matter networks of {M}ajorana zero modes},\ }\href {https://doi.org/10.1038/s41534-021-00508-w} {\bibfield  {journal} {\bibinfo  {journal} {npj Quantum Information}\ }\textbf {\bibinfo {volume} {7}},\ \bibinfo {pages} {171} (\bibinfo {year} {2021})}\BibitemShut {NoStop}%
\bibitem [{\citenamefont {Wang}\ \emph {et~al.}(2019)\citenamefont {Wang}, \citenamefont {Ronca},\ and\ \citenamefont {Sentef}}]{ChernGraphene_Wang}%
  \BibitemOpen
  \bibfield  {author} {\bibinfo {author} {\bibfnamefont {X.}~\bibnamefont {Wang}}, \bibinfo {author} {\bibfnamefont {E.}~\bibnamefont {Ronca}},\ and\ \bibinfo {author} {\bibfnamefont {M.~A.}\ \bibnamefont {Sentef}},\ }\bibfield  {title} {\bibinfo {title} {Cavity quantum electrodynamical {C}hern insulator: {T}owards light-induced quantized anomalous {H}all effect in graphene},\ }\href {https://doi.org/10.1103/PhysRevB.99.235156} {\bibfield  {journal} {\bibinfo  {journal} {Phys. Rev. B}\ }\textbf {\bibinfo {volume} {99}},\ \bibinfo {pages} {235156} (\bibinfo {year} {2019})}\BibitemShut {NoStop}%
\bibitem [{\citenamefont {Li}\ \emph {et~al.}(2022)\citenamefont {Li}, \citenamefont {Schamri\ss{}},\ and\ \citenamefont {Eckstein}}]{li2022effective}%
  \BibitemOpen
  \bibfield  {author} {\bibinfo {author} {\bibfnamefont {J.}~\bibnamefont {Li}}, \bibinfo {author} {\bibfnamefont {L.}~\bibnamefont {Schamri\ss{}}},\ and\ \bibinfo {author} {\bibfnamefont {M.}~\bibnamefont {Eckstein}},\ }\bibfield  {title} {\bibinfo {title} {Effective theory of lattice electrons strongly coupled to quantum electromagnetic fields},\ }\href {https://doi.org/10.1103/PhysRevB.105.165121} {\bibfield  {journal} {\bibinfo  {journal} {Phys. Rev. B}\ }\textbf {\bibinfo {volume} {105}},\ \bibinfo {pages} {165121} (\bibinfo {year} {2022})}\BibitemShut {NoStop}%
\bibitem [{\citenamefont {Dag}\ and\ \citenamefont {Rokaj}(2024)}]{dag2024}%
  \BibitemOpen
  \bibfield  {author} {\bibinfo {author} {\bibfnamefont {C.~B.}\ \bibnamefont {Dag}}\ and\ \bibinfo {author} {\bibfnamefont {V.}~\bibnamefont {Rokaj}},\ }\bibfield  {title} {\bibinfo {title} {Engineering topology in graphene with chiral cavities},\ }\href {https://doi.org/10.1103/PhysRevB.110.L121101} {\bibfield  {journal} {\bibinfo  {journal} {Phys. Rev. B}\ }\textbf {\bibinfo {volume} {110}},\ \bibinfo {pages} {L121101} (\bibinfo {year} {2024})}\BibitemShut {NoStop}%
\bibitem [{\citenamefont {Ghorashi}\ \emph {et~al.}(2025)\citenamefont {Ghorashi}, \citenamefont {Cano},\ and\ \citenamefont {Dag}}]{ghorashi2025tunable}%
  \BibitemOpen
  \bibfield  {author} {\bibinfo {author} {\bibfnamefont {S.~A.~A.}\ \bibnamefont {Ghorashi}}, \bibinfo {author} {\bibfnamefont {J.}~\bibnamefont {Cano}},\ and\ \bibinfo {author} {\bibfnamefont {C.~B.}\ \bibnamefont {Dag}},\ }\href {https://arxiv.org/abs/2504.03842} {\bibinfo {title} {Tunable topological phases in multilayer graphene coupled to a chiral cavity}} (\bibinfo {year} {2025}),\ \Eprint {https://arxiv.org/abs/2504.03842} {arXiv:2504.03842 [cond-mat.mes-hall]} \BibitemShut {NoStop}%
\bibitem [{\citenamefont {Tay}\ \emph {et~al.}(2025)\citenamefont {Tay}, \citenamefont {Sanders}, \citenamefont {Baydin}, \citenamefont {Song}, \citenamefont {Welakuh}, \citenamefont {Alabastri}, \citenamefont {Rokaj}, \citenamefont {Dag},\ and\ \citenamefont {Kono}}]{tay2025terahertz}%
  \BibitemOpen
  \bibfield  {author} {\bibinfo {author} {\bibfnamefont {F.}~\bibnamefont {Tay}}, \bibinfo {author} {\bibfnamefont {S.}~\bibnamefont {Sanders}}, \bibinfo {author} {\bibfnamefont {A.}~\bibnamefont {Baydin}}, \bibinfo {author} {\bibfnamefont {Z.}~\bibnamefont {Song}}, \bibinfo {author} {\bibfnamefont {D.~M.}\ \bibnamefont {Welakuh}}, \bibinfo {author} {\bibfnamefont {A.}~\bibnamefont {Alabastri}}, \bibinfo {author} {\bibfnamefont {V.}~\bibnamefont {Rokaj}}, \bibinfo {author} {\bibfnamefont {C.~B.}\ \bibnamefont {Dag}},\ and\ \bibinfo {author} {\bibfnamefont {J.}~\bibnamefont {Kono}},\ }\bibfield  {title} {\bibinfo {title} {Terahertz chiral photonic-crystal cavities for {D}irac gap engineering in graphene},\ }\href {https://doi.org/10.1038/s41467-025-60335-x} {\bibfield  {journal} {\bibinfo  {journal} {Nature Communications}\ }\textbf {\bibinfo {volume} {16}},\ \bibinfo {pages} {5270} (\bibinfo {year} {2025})}\BibitemShut {NoStop}%
\bibitem [{\citenamefont {Sau}\ and\ \citenamefont {Sarma}(2012)}]{sau2012realizing}%
  \BibitemOpen
  \bibfield  {author} {\bibinfo {author} {\bibfnamefont {J.~D.}\ \bibnamefont {Sau}}\ and\ \bibinfo {author} {\bibfnamefont {S.~D.}\ \bibnamefont {Sarma}},\ }\bibfield  {title} {\bibinfo {title} {Realizing a robust practical {M}ajorana chain in a quantum-dot-superconductor linear array},\ }\href {https://doi.org/10.1038/ncomms1966} {\bibfield  {journal} {\bibinfo  {journal} {Nature Communications}\ }\textbf {\bibinfo {volume} {3}},\ \bibinfo {pages} {964} (\bibinfo {year} {2012})}\BibitemShut {NoStop}%
\bibitem [{\citenamefont {Leijnse}\ and\ \citenamefont {Flensberg}(2012)}]{leijnse2012parity}%
  \BibitemOpen
  \bibfield  {author} {\bibinfo {author} {\bibfnamefont {M.}~\bibnamefont {Leijnse}}\ and\ \bibinfo {author} {\bibfnamefont {K.}~\bibnamefont {Flensberg}},\ }\bibfield  {title} {\bibinfo {title} {Parity qubits and poor man's {M}ajorana bound states in double quantum dots},\ }\href {https://doi.org/10.1103/PhysRevB.86.134528} {\bibfield  {journal} {\bibinfo  {journal} {Phys. Rev. B}\ }\textbf {\bibinfo {volume} {86}},\ \bibinfo {pages} {134528} (\bibinfo {year} {2012})}\BibitemShut {NoStop}%
\bibitem [{\citenamefont {Fulga}\ \emph {et~al.}(2013)\citenamefont {Fulga}, \citenamefont {Haim}, \citenamefont {Akhmerov},\ and\ \citenamefont {Oreg}}]{fulga2013adaptive}%
  \BibitemOpen
  \bibfield  {author} {\bibinfo {author} {\bibfnamefont {I.~C.}\ \bibnamefont {Fulga}}, \bibinfo {author} {\bibfnamefont {A.}~\bibnamefont {Haim}}, \bibinfo {author} {\bibfnamefont {A.~R.}\ \bibnamefont {Akhmerov}},\ and\ \bibinfo {author} {\bibfnamefont {Y.}~\bibnamefont {Oreg}},\ }\bibfield  {title} {\bibinfo {title} {Adaptive tuning of {M}ajorana fermions in a quantum dot chain},\ }\href {https://doi.org/10.1088/1367-2630/15/4/045020} {\bibfield  {journal} {\bibinfo  {journal} {New Journal of Physics}\ }\textbf {\bibinfo {volume} {15}},\ \bibinfo {pages} {045020} (\bibinfo {year} {2013})}\BibitemShut {NoStop}%
\bibitem [{\citenamefont {Seoane~Souto}\ and\ \citenamefont {Aguado}(2024)}]{SeoaneSouto2024}%
  \BibitemOpen
  \bibfield  {author} {\bibinfo {author} {\bibfnamefont {R.}~\bibnamefont {Seoane~Souto}}\ and\ \bibinfo {author} {\bibfnamefont {R.}~\bibnamefont {Aguado}},\ }\bibinfo {title} {Subgap states in semiconductor-superconductor devices for quantum technologies: {A}ndreev qubits and minimal {M}ajorana chains},\ in\ \href {https://doi.org/10.1007/978-3-031-55657-9_3} {\emph {\bibinfo {booktitle} {New Trends and Platforms for Quantum Technologies}}},\ \bibinfo {editor} {edited by\ \bibinfo {editor} {\bibfnamefont {R.}~\bibnamefont {Aguado}}, \bibinfo {editor} {\bibfnamefont {R.}~\bibnamefont {Citro}}, \bibinfo {editor} {\bibfnamefont {M.}~\bibnamefont {Lewenstein}},\ and\ \bibinfo {editor} {\bibfnamefont {M.}~\bibnamefont {Stern}}}\ (\bibinfo  {publisher} {Springer Nature Switzerland},\ \bibinfo {address} {Cham},\ \bibinfo {year} {2024})\ pp.\ \bibinfo {pages} {133--223}\BibitemShut {NoStop}%
\bibitem [{\citenamefont {Dvir}\ \emph {et~al.}(2023)\citenamefont {Dvir}, \citenamefont {Wang}, \citenamefont {van Loo}, \citenamefont {Liu}, \citenamefont {Mazur}, \citenamefont {Bordin}, \citenamefont {ten Haaf}, \citenamefont {Wang}, \citenamefont {van Driel}, \citenamefont {Zatelli}, \citenamefont {Li}, \citenamefont {Malinowski}, \citenamefont {Gazibegovic}, \citenamefont {Badawy}, \citenamefont {Bakkers}, \citenamefont {Wimmer},\ and\ \citenamefont {Kouwenhoven}}]{dvir2023realization}%
  \BibitemOpen
  \bibfield  {author} {\bibinfo {author} {\bibfnamefont {T.}~\bibnamefont {Dvir}}, \bibinfo {author} {\bibfnamefont {G.}~\bibnamefont {Wang}}, \bibinfo {author} {\bibfnamefont {N.}~\bibnamefont {van Loo}}, \bibinfo {author} {\bibfnamefont {C.-X.}\ \bibnamefont {Liu}}, \bibinfo {author} {\bibfnamefont {G.~P.}\ \bibnamefont {Mazur}}, \bibinfo {author} {\bibfnamefont {A.}~\bibnamefont {Bordin}}, \bibinfo {author} {\bibfnamefont {S.~L.~D.}\ \bibnamefont {ten Haaf}}, \bibinfo {author} {\bibfnamefont {J.-Y.}\ \bibnamefont {Wang}}, \bibinfo {author} {\bibfnamefont {D.}~\bibnamefont {van Driel}}, \bibinfo {author} {\bibfnamefont {F.}~\bibnamefont {Zatelli}}, \bibinfo {author} {\bibfnamefont {X.}~\bibnamefont {Li}}, \bibinfo {author} {\bibfnamefont {F.~K.}\ \bibnamefont {Malinowski}}, \bibinfo {author} {\bibfnamefont {S.}~\bibnamefont {Gazibegovic}}, \bibinfo {author} {\bibfnamefont {G.}~\bibnamefont {Badawy}}, \bibinfo {author} {\bibfnamefont {E.~P. A.~M.}\ \bibnamefont {Bakkers}}, \bibinfo {author} {\bibfnamefont
  {M.}~\bibnamefont {Wimmer}},\ and\ \bibinfo {author} {\bibfnamefont {L.~P.}\ \bibnamefont {Kouwenhoven}},\ }\bibfield  {title} {\bibinfo {title} {Realization of a minimal {K}itaev chain in coupled quantum dots},\ }\href {https://doi.org/10.1038/s41586-022-05585-1} {\bibfield  {journal} {\bibinfo  {journal} {Nature}\ }\textbf {\bibinfo {volume} {614}},\ \bibinfo {pages} {445} (\bibinfo {year} {2023})}\BibitemShut {NoStop}%
\bibitem [{\citenamefont {Zatelli}\ \emph {et~al.}(2024)\citenamefont {Zatelli}, \citenamefont {van Driel}, \citenamefont {Xu}, \citenamefont {Wang}, \citenamefont {Liu}, \citenamefont {Bordin}, \citenamefont {Roovers}, \citenamefont {Mazur}, \citenamefont {van Loo}, \citenamefont {Wolff}, \citenamefont {Bozkurt}, \citenamefont {Badawy}, \citenamefont {Gazibegovic}, \citenamefont {Bakkers}, \citenamefont {Wimmer}, \citenamefont {Kouwenhoven},\ and\ \citenamefont {Dvir}}]{zatelli2023robust}%
  \BibitemOpen
  \bibfield  {author} {\bibinfo {author} {\bibfnamefont {F.}~\bibnamefont {Zatelli}}, \bibinfo {author} {\bibfnamefont {D.}~\bibnamefont {van Driel}}, \bibinfo {author} {\bibfnamefont {D.}~\bibnamefont {Xu}}, \bibinfo {author} {\bibfnamefont {G.}~\bibnamefont {Wang}}, \bibinfo {author} {\bibfnamefont {C.-X.}\ \bibnamefont {Liu}}, \bibinfo {author} {\bibfnamefont {A.}~\bibnamefont {Bordin}}, \bibinfo {author} {\bibfnamefont {B.}~\bibnamefont {Roovers}}, \bibinfo {author} {\bibfnamefont {G.~P.}\ \bibnamefont {Mazur}}, \bibinfo {author} {\bibfnamefont {N.}~\bibnamefont {van Loo}}, \bibinfo {author} {\bibfnamefont {J.~C.}\ \bibnamefont {Wolff}}, \bibinfo {author} {\bibfnamefont {A.~M.}\ \bibnamefont {Bozkurt}}, \bibinfo {author} {\bibfnamefont {G.}~\bibnamefont {Badawy}}, \bibinfo {author} {\bibfnamefont {S.}~\bibnamefont {Gazibegovic}}, \bibinfo {author} {\bibfnamefont {E.~P. A.~M.}\ \bibnamefont {Bakkers}}, \bibinfo {author} {\bibfnamefont {M.}~\bibnamefont {Wimmer}}, \bibinfo {author} {\bibfnamefont {L.~P.}\
  \bibnamefont {Kouwenhoven}},\ and\ \bibinfo {author} {\bibfnamefont {T.}~\bibnamefont {Dvir}},\ }\bibfield  {title} {\bibinfo {title} {Robust poor man's {M}ajorana zero modes using {Y}u-{S}hiba-{R}usinov states},\ }\href {https://doi.org/10.1038/s41467-024-52066-2} {\bibfield  {journal} {\bibinfo  {journal} {Nature Communications}\ }\textbf {\bibinfo {volume} {15}},\ \bibinfo {pages} {7933} (\bibinfo {year} {2024})}\BibitemShut {NoStop}%
\bibitem [{\citenamefont {ten Haaf}\ \emph {et~al.}(2024)\citenamefont {ten Haaf}, \citenamefont {Wang}, \citenamefont {Bozkurt}, \citenamefont {Liu}, \citenamefont {Kulesh}, \citenamefont {Kim}, \citenamefont {Xiao}, \citenamefont {Thomas}, \citenamefont {Manfra}, \citenamefont {Dvir}, \citenamefont {Wimmer},\ and\ \citenamefont {Goswami}}]{tenhaaf2023engineering}%
  \BibitemOpen
  \bibfield  {author} {\bibinfo {author} {\bibfnamefont {S.~L.~D.}\ \bibnamefont {ten Haaf}}, \bibinfo {author} {\bibfnamefont {Q.}~\bibnamefont {Wang}}, \bibinfo {author} {\bibfnamefont {A.~M.}\ \bibnamefont {Bozkurt}}, \bibinfo {author} {\bibfnamefont {C.-X.}\ \bibnamefont {Liu}}, \bibinfo {author} {\bibfnamefont {I.}~\bibnamefont {Kulesh}}, \bibinfo {author} {\bibfnamefont {P.}~\bibnamefont {Kim}}, \bibinfo {author} {\bibfnamefont {D.}~\bibnamefont {Xiao}}, \bibinfo {author} {\bibfnamefont {C.}~\bibnamefont {Thomas}}, \bibinfo {author} {\bibfnamefont {M.~J.}\ \bibnamefont {Manfra}}, \bibinfo {author} {\bibfnamefont {T.}~\bibnamefont {Dvir}}, \bibinfo {author} {\bibfnamefont {M.}~\bibnamefont {Wimmer}},\ and\ \bibinfo {author} {\bibfnamefont {S.}~\bibnamefont {Goswami}},\ }\bibfield  {title} {\bibinfo {title} {A two-site {K}itaev chain in a two-dimensional electron gas},\ }\href {https://doi.org/10.1038/s41586-024-07434-9} {\bibfield  {journal} {\bibinfo  {journal} {Nature}\ }\textbf {\bibinfo {volume}
  {630}},\ \bibinfo {pages} {329} (\bibinfo {year} {2024})}\BibitemShut {NoStop}%
\bibitem [{\citenamefont {Bordin}\ \emph {et~al.}(2025)\citenamefont {Bordin}, \citenamefont {Liu}, \citenamefont {Dvir}, \citenamefont {Zatelli}, \citenamefont {ten Haaf}, \citenamefont {van Driel}, \citenamefont {Wang}, \citenamefont {van Loo}, \citenamefont {Zhang}, \citenamefont {Wolff}, \citenamefont {Van~Caekenberghe}, \citenamefont {Badawy}, \citenamefont {Gazibegovic}, \citenamefont {Bakkers}, \citenamefont {Wimmer}, \citenamefont {Kouwenhoven},\ and\ \citenamefont {Mazur}}]{bordin2024signatures}%
  \BibitemOpen
  \bibfield  {author} {\bibinfo {author} {\bibfnamefont {A.}~\bibnamefont {Bordin}}, \bibinfo {author} {\bibfnamefont {C.-X.}\ \bibnamefont {Liu}}, \bibinfo {author} {\bibfnamefont {T.}~\bibnamefont {Dvir}}, \bibinfo {author} {\bibfnamefont {F.}~\bibnamefont {Zatelli}}, \bibinfo {author} {\bibfnamefont {S.~L.~D.}\ \bibnamefont {ten Haaf}}, \bibinfo {author} {\bibfnamefont {D.}~\bibnamefont {van Driel}}, \bibinfo {author} {\bibfnamefont {G.}~\bibnamefont {Wang}}, \bibinfo {author} {\bibfnamefont {N.}~\bibnamefont {van Loo}}, \bibinfo {author} {\bibfnamefont {Y.}~\bibnamefont {Zhang}}, \bibinfo {author} {\bibfnamefont {J.~C.}\ \bibnamefont {Wolff}}, \bibinfo {author} {\bibfnamefont {T.}~\bibnamefont {Van~Caekenberghe}}, \bibinfo {author} {\bibfnamefont {G.}~\bibnamefont {Badawy}}, \bibinfo {author} {\bibfnamefont {S.}~\bibnamefont {Gazibegovic}}, \bibinfo {author} {\bibfnamefont {E.~P. A.~M.}\ \bibnamefont {Bakkers}}, \bibinfo {author} {\bibfnamefont {M.}~\bibnamefont {Wimmer}}, \bibinfo {author} {\bibfnamefont
  {L.~P.}\ \bibnamefont {Kouwenhoven}},\ and\ \bibinfo {author} {\bibfnamefont {G.~P.}\ \bibnamefont {Mazur}},\ }\bibfield  {title} {\bibinfo {title} {Enhanced {M}ajorana stability in a three-site {K}itaev chain},\ }\bibfield  {journal} {\bibinfo  {journal} {Nature Nanotechnology}\ }\href {https://doi.org/10.1038/s41565-025-01894-4} {10.1038/s41565-025-01894-4} (\bibinfo {year} {2025})\BibitemShut {NoStop}%
\bibitem [{\citenamefont {Bruhat}\ \emph {et~al.}(2018)\citenamefont {Bruhat}, \citenamefont {Cubaynes}, \citenamefont {Viennot}, \citenamefont {Dartiailh}, \citenamefont {Desjardins}, \citenamefont {Cottet},\ and\ \citenamefont {Kontos}}]{CavDQDsuper2018}%
  \BibitemOpen
  \bibfield  {author} {\bibinfo {author} {\bibfnamefont {L.~E.}\ \bibnamefont {Bruhat}}, \bibinfo {author} {\bibfnamefont {T.}~\bibnamefont {Cubaynes}}, \bibinfo {author} {\bibfnamefont {J.~J.}\ \bibnamefont {Viennot}}, \bibinfo {author} {\bibfnamefont {M.~C.}\ \bibnamefont {Dartiailh}}, \bibinfo {author} {\bibfnamefont {M.~M.}\ \bibnamefont {Desjardins}}, \bibinfo {author} {\bibfnamefont {A.}~\bibnamefont {Cottet}},\ and\ \bibinfo {author} {\bibfnamefont {T.}~\bibnamefont {Kontos}},\ }\bibfield  {title} {\bibinfo {title} {Circuit {QED} with a quantum-dot charge qubit dressed by {C}ooper pairs},\ }\href {https://doi.org/10.1103/PhysRevB.98.155313} {\bibfield  {journal} {\bibinfo  {journal} {Phys. Rev. B}\ }\textbf {\bibinfo {volume} {98}},\ \bibinfo {pages} {155313} (\bibinfo {year} {2018})}\BibitemShut {NoStop}%
\bibitem [{\citenamefont {Scarlino}\ \emph {et~al.}(2019)\citenamefont {Scarlino}, \citenamefont {van Woerkom}, \citenamefont {Mendes}, \citenamefont {Koski}, \citenamefont {Landig}, \citenamefont {Andersen}, \citenamefont {Gasparinetti}, \citenamefont {Reichl}, \citenamefont {Wegscheider}, \citenamefont {Ensslin}, \citenamefont {Ihn}, \citenamefont {Blais},\ and\ \citenamefont {Wallraff}}]{CavDQDsuper2019}%
  \BibitemOpen
  \bibfield  {author} {\bibinfo {author} {\bibfnamefont {P.}~\bibnamefont {Scarlino}}, \bibinfo {author} {\bibfnamefont {D.~J.}\ \bibnamefont {van Woerkom}}, \bibinfo {author} {\bibfnamefont {U.~C.}\ \bibnamefont {Mendes}}, \bibinfo {author} {\bibfnamefont {J.~V.}\ \bibnamefont {Koski}}, \bibinfo {author} {\bibfnamefont {A.~J.}\ \bibnamefont {Landig}}, \bibinfo {author} {\bibfnamefont {C.~K.}\ \bibnamefont {Andersen}}, \bibinfo {author} {\bibfnamefont {S.}~\bibnamefont {Gasparinetti}}, \bibinfo {author} {\bibfnamefont {C.}~\bibnamefont {Reichl}}, \bibinfo {author} {\bibfnamefont {W.}~\bibnamefont {Wegscheider}}, \bibinfo {author} {\bibfnamefont {K.}~\bibnamefont {Ensslin}}, \bibinfo {author} {\bibfnamefont {T.}~\bibnamefont {Ihn}}, \bibinfo {author} {\bibfnamefont {A.}~\bibnamefont {Blais}},\ and\ \bibinfo {author} {\bibfnamefont {A.}~\bibnamefont {Wallraff}},\ }\bibfield  {title} {\bibinfo {title} {Coherent microwave-photon-mediated coupling between a semiconductor and a superconducting qubit},\ }\href
  {https://doi.org/10.1038/s41467-019-10798-6} {\bibfield  {journal} {\bibinfo  {journal} {Nature Communications}\ }\textbf {\bibinfo {volume} {10}},\ \bibinfo {pages} {3011} (\bibinfo {year} {2019})}\BibitemShut {NoStop}%
\bibitem [{\citenamefont {G\'omez-Le\'on}\ \emph {et~al.}(2025)\citenamefont {G\'omez-Le\'on}, \citenamefont {Schir\`o},\ and\ \citenamefont {Dmytruk}}]{gomez2024high}%
  \BibitemOpen
  \bibfield  {author} {\bibinfo {author} {\bibfnamefont {A.}~\bibnamefont {G\'omez-Le\'on}}, \bibinfo {author} {\bibfnamefont {M.}~\bibnamefont {Schir\`o}},\ and\ \bibinfo {author} {\bibfnamefont {O.}~\bibnamefont {Dmytruk}},\ }\bibfield  {title} {\bibinfo {title} {Majorana bound states from cavity embedding in an interacting two-site {K}itaev chain},\ }\href {https://doi.org/10.1103/PhysRevB.111.155410} {\bibfield  {journal} {\bibinfo  {journal} {Phys. Rev. B}\ }\textbf {\bibinfo {volume} {111}},\ \bibinfo {pages} {155410} (\bibinfo {year} {2025})}\BibitemShut {NoStop}%
\bibitem [{\citenamefont {Passetti}\ \emph {et~al.}(2023)\citenamefont {Passetti}, \citenamefont {Eckhardt}, \citenamefont {Sentef},\ and\ \citenamefont {Kennes}}]{passetti2023cavity}%
  \BibitemOpen
  \bibfield  {author} {\bibinfo {author} {\bibfnamefont {G.}~\bibnamefont {Passetti}}, \bibinfo {author} {\bibfnamefont {C.~J.}\ \bibnamefont {Eckhardt}}, \bibinfo {author} {\bibfnamefont {M.~A.}\ \bibnamefont {Sentef}},\ and\ \bibinfo {author} {\bibfnamefont {D.~M.}\ \bibnamefont {Kennes}},\ }\bibfield  {title} {\bibinfo {title} {Cavity light-matter entanglement through quantum fluctuations},\ }\href {https://doi.org/10.1103/PhysRevLett.131.023601} {\bibfield  {journal} {\bibinfo  {journal} {Phys. Rev. Lett.}\ }\textbf {\bibinfo {volume} {131}},\ \bibinfo {pages} {023601} (\bibinfo {year} {2023})}\BibitemShut {NoStop}%
\bibitem [{\citenamefont {Li}\ and\ \citenamefont {Eckstein}(2020)}]{Li_ED_Hubbard}%
  \BibitemOpen
  \bibfield  {author} {\bibinfo {author} {\bibfnamefont {J.}~\bibnamefont {Li}}\ and\ \bibinfo {author} {\bibfnamefont {M.}~\bibnamefont {Eckstein}},\ }\bibfield  {title} {\bibinfo {title} {Manipulating intertwined orders in solids with quantum light},\ }\href {https://doi.org/10.1103/PhysRevLett.125.217402} {\bibfield  {journal} {\bibinfo  {journal} {Phys. Rev. Lett.}\ }\textbf {\bibinfo {volume} {125}},\ \bibinfo {pages} {217402} (\bibinfo {year} {2020})}\BibitemShut {NoStop}%
\bibitem [{\citenamefont {Nakamoto}\ \emph {et~al.}(2025)\citenamefont {Nakamoto}, \citenamefont {Takasan},\ and\ \citenamefont {Tsuji}}]{Hubbard_Cav_Nakamoto}%
  \BibitemOpen
  \bibfield  {author} {\bibinfo {author} {\bibfnamefont {T.}~\bibnamefont {Nakamoto}}, \bibinfo {author} {\bibfnamefont {K.}~\bibnamefont {Takasan}},\ and\ \bibinfo {author} {\bibfnamefont {N.}~\bibnamefont {Tsuji}},\ }\href {https://arxiv.org/abs/2505.09311} {\bibinfo {title} {One-dimensional extended {H}ubbard model coupled with an optical cavity}} (\bibinfo {year} {2025}),\ \Eprint {https://arxiv.org/abs/2505.09311} {arXiv:2505.09311 [cond-mat.str-el]} \BibitemShut {NoStop}%
\bibitem [{\citenamefont {Kass}\ \emph {et~al.}(2024)\citenamefont {Kass}, \citenamefont {Talkington}, \citenamefont {Srivastava},\ and\ \citenamefont {Claassen}}]{kass2024manybodyphotonblockadequantum}%
  \BibitemOpen
  \bibfield  {author} {\bibinfo {author} {\bibfnamefont {B.}~\bibnamefont {Kass}}, \bibinfo {author} {\bibfnamefont {S.}~\bibnamefont {Talkington}}, \bibinfo {author} {\bibfnamefont {A.}~\bibnamefont {Srivastava}},\ and\ \bibinfo {author} {\bibfnamefont {M.}~\bibnamefont {Claassen}},\ }\href {https://arxiv.org/abs/2411.08964} {\bibinfo {title} {Many-body photon blockade and quantum light generation from cavity quantum materials}} (\bibinfo {year} {2024}),\ \Eprint {https://arxiv.org/abs/2411.08964} {arXiv:2411.08964 [cond-mat.str-el]} \BibitemShut {NoStop}%
\bibitem [{\citenamefont {M\'endez-C\'ordoba}\ \emph {et~al.}(2020)\citenamefont {M\'endez-C\'ordoba}, \citenamefont {Mendoza-Arenas}, \citenamefont {G\'omez-Ruiz}, \citenamefont {Rodr\'{\i}guez}, \citenamefont {Tejedor},\ and\ \citenamefont {Quiroga}}]{PhotonNumber2020}%
  \BibitemOpen
  \bibfield  {author} {\bibinfo {author} {\bibfnamefont {F.~P.~M.}\ \bibnamefont {M\'endez-C\'ordoba}}, \bibinfo {author} {\bibfnamefont {J.~J.}\ \bibnamefont {Mendoza-Arenas}}, \bibinfo {author} {\bibfnamefont {F.~J.}\ \bibnamefont {G\'omez-Ruiz}}, \bibinfo {author} {\bibfnamefont {F.~J.}\ \bibnamefont {Rodr\'{\i}guez}}, \bibinfo {author} {\bibfnamefont {C.}~\bibnamefont {Tejedor}},\ and\ \bibinfo {author} {\bibfnamefont {L.}~\bibnamefont {Quiroga}},\ }\bibfield  {title} {\bibinfo {title} {R\'enyi entropy singularities as signatures of topological criticality in coupled photon-fermion systems},\ }\href {https://doi.org/10.1103/PhysRevResearch.2.043264} {\bibfield  {journal} {\bibinfo  {journal} {Phys. Rev. Res.}\ }\textbf {\bibinfo {volume} {2}},\ \bibinfo {pages} {043264} (\bibinfo {year} {2020})}\BibitemShut {NoStop}%
\bibitem [{\citenamefont {Lysne}\ \emph {et~al.}(2023)\citenamefont {Lysne}, \citenamefont {Sch\"uler},\ and\ \citenamefont {Werner}}]{PhotElec_measure_Lysne}%
  \BibitemOpen
  \bibfield  {author} {\bibinfo {author} {\bibfnamefont {M.}~\bibnamefont {Lysne}}, \bibinfo {author} {\bibfnamefont {M.}~\bibnamefont {Sch\"uler}},\ and\ \bibinfo {author} {\bibfnamefont {P.}~\bibnamefont {Werner}},\ }\bibfield  {title} {\bibinfo {title} {Quantum optics measurement scheme for quantum geometry and topological invariants},\ }\href {https://doi.org/10.1103/PhysRevLett.131.156901} {\bibfield  {journal} {\bibinfo  {journal} {Phys. Rev. Lett.}\ }\textbf {\bibinfo {volume} {131}},\ \bibinfo {pages} {156901} (\bibinfo {year} {2023})}\BibitemShut {NoStop}%
\bibitem [{\citenamefont {Grunwald}\ \emph {et~al.}(2024)\citenamefont {Grunwald}, \citenamefont {Boström}, \citenamefont {Svendsen}, \citenamefont {Kennes},\ and\ \citenamefont {Rubio}}]{grunwald2024cavity}%
  \BibitemOpen
  \bibfield  {author} {\bibinfo {author} {\bibfnamefont {L.}~\bibnamefont {Grunwald}}, \bibinfo {author} {\bibfnamefont {E.~V.}\ \bibnamefont {Boström}}, \bibinfo {author} {\bibfnamefont {M.~K.}\ \bibnamefont {Svendsen}}, \bibinfo {author} {\bibfnamefont {D.~M.}\ \bibnamefont {Kennes}},\ and\ \bibinfo {author} {\bibfnamefont {A.}~\bibnamefont {Rubio}},\ }\href {https://arxiv.org/abs/2410.21515} {\bibinfo {title} {Cavity spectroscopy for strongly correlated systems}} (\bibinfo {year} {2024}),\ \Eprint {https://arxiv.org/abs/2410.21515} {arXiv:2410.21515 [cond-mat.str-el]} \BibitemShut {NoStop}%
\bibitem [{\citenamefont {Schulte}\ \emph {et~al.}(2015)\citenamefont {Schulte}, \citenamefont {Hansom}, \citenamefont {Jones}, \citenamefont {Matthiesen}, \citenamefont {Le~Gall},\ and\ \citenamefont {Atat{\"u}re}}]{QDsqueeze_Schulte}%
  \BibitemOpen
  \bibfield  {author} {\bibinfo {author} {\bibfnamefont {C.~H.~H.}\ \bibnamefont {Schulte}}, \bibinfo {author} {\bibfnamefont {J.}~\bibnamefont {Hansom}}, \bibinfo {author} {\bibfnamefont {A.~E.}\ \bibnamefont {Jones}}, \bibinfo {author} {\bibfnamefont {C.}~\bibnamefont {Matthiesen}}, \bibinfo {author} {\bibfnamefont {C.}~\bibnamefont {Le~Gall}},\ and\ \bibinfo {author} {\bibfnamefont {M.}~\bibnamefont {Atat{\"u}re}},\ }\bibfield  {title} {\bibinfo {title} {Quadrature squeezed photons from a two-level system},\ }\href {https://doi.org/10.1038/nature14868} {\bibfield  {journal} {\bibinfo  {journal} {Nature}\ }\textbf {\bibinfo {volume} {525}},\ \bibinfo {pages} {222} (\bibinfo {year} {2015})}\BibitemShut {NoStop}%
\bibitem [{\citenamefont {Brooks}\ \emph {et~al.}(2012)\citenamefont {Brooks}, \citenamefont {Botter}, \citenamefont {Schreppler}, \citenamefont {Purdy}, \citenamefont {Brahms},\ and\ \citenamefont {Stamper-Kurn}}]{optomech_Brooks}%
  \BibitemOpen
  \bibfield  {author} {\bibinfo {author} {\bibfnamefont {D.~W.~C.}\ \bibnamefont {Brooks}}, \bibinfo {author} {\bibfnamefont {T.}~\bibnamefont {Botter}}, \bibinfo {author} {\bibfnamefont {S.}~\bibnamefont {Schreppler}}, \bibinfo {author} {\bibfnamefont {T.~P.}\ \bibnamefont {Purdy}}, \bibinfo {author} {\bibfnamefont {N.}~\bibnamefont {Brahms}},\ and\ \bibinfo {author} {\bibfnamefont {D.~M.}\ \bibnamefont {Stamper-Kurn}},\ }\bibfield  {title} {\bibinfo {title} {Non-classical light generated by quantum-noise-driven cavity optomechanics},\ }\href {https://doi.org/10.1038/nature11325} {\bibfield  {journal} {\bibinfo  {journal} {Nature}\ }\textbf {\bibinfo {volume} {488}},\ \bibinfo {pages} {476} (\bibinfo {year} {2012})}\BibitemShut {NoStop}%
\bibitem [{\citenamefont {Safavi-Naeini}\ \emph {et~al.}(2013)\citenamefont {Safavi-Naeini}, \citenamefont {Gr{\"o}blacher}, \citenamefont {Hill}, \citenamefont {Chan}, \citenamefont {Aspelmeyer},\ and\ \citenamefont {Painter}}]{optomech_Safavi-Naeini}%
  \BibitemOpen
  \bibfield  {author} {\bibinfo {author} {\bibfnamefont {A.~H.}\ \bibnamefont {Safavi-Naeini}}, \bibinfo {author} {\bibfnamefont {S.}~\bibnamefont {Gr{\"o}blacher}}, \bibinfo {author} {\bibfnamefont {J.~T.}\ \bibnamefont {Hill}}, \bibinfo {author} {\bibfnamefont {J.}~\bibnamefont {Chan}}, \bibinfo {author} {\bibfnamefont {M.}~\bibnamefont {Aspelmeyer}},\ and\ \bibinfo {author} {\bibfnamefont {O.}~\bibnamefont {Painter}},\ }\bibfield  {title} {\bibinfo {title} {Squeezed light from a silicon micromechanical resonator},\ }\href {https://doi.org/10.1038/nature12307} {\bibfield  {journal} {\bibinfo  {journal} {Nature}\ }\textbf {\bibinfo {volume} {500}},\ \bibinfo {pages} {185} (\bibinfo {year} {2013})}\BibitemShut {NoStop}%
\bibitem [{\citenamefont {Walls}\ and\ \citenamefont {Zoller}(1981)}]{squeeze_predict}%
  \BibitemOpen
  \bibfield  {author} {\bibinfo {author} {\bibfnamefont {D.~F.}\ \bibnamefont {Walls}}\ and\ \bibinfo {author} {\bibfnamefont {P.}~\bibnamefont {Zoller}},\ }\bibfield  {title} {\bibinfo {title} {Reduced quantum fluctuations in resonance fluorescence},\ }\href {https://doi.org/10.1103/PhysRevLett.47.709} {\bibfield  {journal} {\bibinfo  {journal} {Phys. Rev. Lett.}\ }\textbf {\bibinfo {volume} {47}},\ \bibinfo {pages} {709} (\bibinfo {year} {1981})}\BibitemShut {NoStop}%
\bibitem [{\citenamefont {Andersen}\ \emph {et~al.}(2016)\citenamefont {Andersen}, \citenamefont {Gehring}, \citenamefont {Marquardt},\ and\ \citenamefont {Leuchs}}]{squeezeRev_Andersen}%
  \BibitemOpen
  \bibfield  {author} {\bibinfo {author} {\bibfnamefont {U.~L.}\ \bibnamefont {Andersen}}, \bibinfo {author} {\bibfnamefont {T.}~\bibnamefont {Gehring}}, \bibinfo {author} {\bibfnamefont {C.}~\bibnamefont {Marquardt}},\ and\ \bibinfo {author} {\bibfnamefont {G.}~\bibnamefont {Leuchs}},\ }\bibfield  {title} {\bibinfo {title} {30 years of squeezed light generation},\ }\href {https://doi.org/10.1088/0031-8949/91/5/053001} {\bibfield  {journal} {\bibinfo  {journal} {Physica Scripta}\ }\textbf {\bibinfo {volume} {91}},\ \bibinfo {pages} {053001} (\bibinfo {year} {2016})}\BibitemShut {NoStop}%
\bibitem [{\citenamefont {Abadie}\ \emph {et~al.}(2011)\citenamefont {Abadie}, \citenamefont {Abbott}, \citenamefont {Abbott}, \citenamefont {Abbott}, \citenamefont {Abernathy}, \citenamefont {Adams}, \citenamefont {Adhikari}, \citenamefont {Affeldt}, \citenamefont {Allen}, \citenamefont {Allen} \emph {et~al.}}]{GravWaves2011}%
  \BibitemOpen
  \bibfield  {author} {\bibinfo {author} {\bibfnamefont {J.}~\bibnamefont {Abadie}}, \bibinfo {author} {\bibfnamefont {B.~P.}\ \bibnamefont {Abbott}}, \bibinfo {author} {\bibfnamefont {R.}~\bibnamefont {Abbott}}, \bibinfo {author} {\bibfnamefont {T.~D.}\ \bibnamefont {Abbott}}, \bibinfo {author} {\bibfnamefont {M.}~\bibnamefont {Abernathy}}, \bibinfo {author} {\bibfnamefont {C.}~\bibnamefont {Adams}}, \bibinfo {author} {\bibfnamefont {R.}~\bibnamefont {Adhikari}}, \bibinfo {author} {\bibfnamefont {C.}~\bibnamefont {Affeldt}}, \bibinfo {author} {\bibfnamefont {B.}~\bibnamefont {Allen}}, \bibinfo {author} {\bibfnamefont {G.~S.}\ \bibnamefont {Allen}}, \emph {et~al.},\ }\bibfield  {title} {\bibinfo {title} {A gravitational wave observatory operating beyond the quantum shot-noise limit},\ }\href {https://doi.org/10.1038/nphys2083} {\bibfield  {journal} {\bibinfo  {journal} {Nature Physics}\ }\textbf {\bibinfo {volume} {7}},\ \bibinfo {pages} {962} (\bibinfo {year} {2011})}\BibitemShut {NoStop}%
\bibitem [{\citenamefont {Aasi}\ \emph {et~al.}(2013)\citenamefont {Aasi}, \citenamefont {Abadie}, \citenamefont {Abbott}, \citenamefont {Abbott}, \citenamefont {Abbott}, \citenamefont {Abernathy}, \citenamefont {Adams}, \citenamefont {Adams}, \citenamefont {Addesso}, \citenamefont {Adhikari} \emph {et~al.}}]{GravWaves2013}%
  \BibitemOpen
  \bibfield  {author} {\bibinfo {author} {\bibfnamefont {J.}~\bibnamefont {Aasi}}, \bibinfo {author} {\bibfnamefont {J.}~\bibnamefont {Abadie}}, \bibinfo {author} {\bibfnamefont {B.~P.}\ \bibnamefont {Abbott}}, \bibinfo {author} {\bibfnamefont {R.}~\bibnamefont {Abbott}}, \bibinfo {author} {\bibfnamefont {T.~D.}\ \bibnamefont {Abbott}}, \bibinfo {author} {\bibfnamefont {M.~R.}\ \bibnamefont {Abernathy}}, \bibinfo {author} {\bibfnamefont {C.}~\bibnamefont {Adams}}, \bibinfo {author} {\bibfnamefont {T.}~\bibnamefont {Adams}}, \bibinfo {author} {\bibfnamefont {P.}~\bibnamefont {Addesso}}, \bibinfo {author} {\bibfnamefont {R.~X.}\ \bibnamefont {Adhikari}}, \emph {et~al.},\ }\bibfield  {title} {\bibinfo {title} {Enhanced sensitivity of the ligo gravitational wave detector by using squeezed states of light},\ }\href {https://doi.org/10.1038/nphoton.2013.177} {\bibfield  {journal} {\bibinfo  {journal} {Nature Photonics}\ }\textbf {\bibinfo {volume} {7}},\ \bibinfo {pages} {613} (\bibinfo {year} {2013})}\BibitemShut
  {NoStop}%
\bibitem [{\citenamefont {Zhong}\ \emph {et~al.}(2020)\citenamefont {Zhong}, \citenamefont {Wang}, \citenamefont {Deng}, \citenamefont {Chen}, \citenamefont {Peng}, \citenamefont {Luo}, \citenamefont {Qin}, \citenamefont {Wu}, \citenamefont {Ding}, \citenamefont {Hu}, \citenamefont {Hu}, \citenamefont {Yang}, \citenamefont {Zhang}, \citenamefont {Li}, \citenamefont {Li}, \citenamefont {Jiang}, \citenamefont {Gan}, \citenamefont {Yang}, \citenamefont {You}, \citenamefont {Wang}, \citenamefont {Li}, \citenamefont {Liu}, \citenamefont {Lu},\ and\ \citenamefont {Pan}}]{squeeze_QComp_Zhong}%
  \BibitemOpen
  \bibfield  {author} {\bibinfo {author} {\bibfnamefont {H.-S.}\ \bibnamefont {Zhong}}, \bibinfo {author} {\bibfnamefont {H.}~\bibnamefont {Wang}}, \bibinfo {author} {\bibfnamefont {Y.-H.}\ \bibnamefont {Deng}}, \bibinfo {author} {\bibfnamefont {M.-C.}\ \bibnamefont {Chen}}, \bibinfo {author} {\bibfnamefont {L.-C.}\ \bibnamefont {Peng}}, \bibinfo {author} {\bibfnamefont {Y.-H.}\ \bibnamefont {Luo}}, \bibinfo {author} {\bibfnamefont {J.}~\bibnamefont {Qin}}, \bibinfo {author} {\bibfnamefont {D.}~\bibnamefont {Wu}}, \bibinfo {author} {\bibfnamefont {X.}~\bibnamefont {Ding}}, \bibinfo {author} {\bibfnamefont {Y.}~\bibnamefont {Hu}}, \bibinfo {author} {\bibfnamefont {P.}~\bibnamefont {Hu}}, \bibinfo {author} {\bibfnamefont {X.-Y.}\ \bibnamefont {Yang}}, \bibinfo {author} {\bibfnamefont {W.-J.}\ \bibnamefont {Zhang}}, \bibinfo {author} {\bibfnamefont {H.}~\bibnamefont {Li}}, \bibinfo {author} {\bibfnamefont {Y.}~\bibnamefont {Li}}, \bibinfo {author} {\bibfnamefont {X.}~\bibnamefont {Jiang}}, \bibinfo {author}
  {\bibfnamefont {L.}~\bibnamefont {Gan}}, \bibinfo {author} {\bibfnamefont {G.}~\bibnamefont {Yang}}, \bibinfo {author} {\bibfnamefont {L.}~\bibnamefont {You}}, \bibinfo {author} {\bibfnamefont {Z.}~\bibnamefont {Wang}}, \bibinfo {author} {\bibfnamefont {L.}~\bibnamefont {Li}}, \bibinfo {author} {\bibfnamefont {N.-L.}\ \bibnamefont {Liu}}, \bibinfo {author} {\bibfnamefont {C.-Y.}\ \bibnamefont {Lu}},\ and\ \bibinfo {author} {\bibfnamefont {J.-W.}\ \bibnamefont {Pan}},\ }\bibfield  {title} {\bibinfo {title} {Quantum computational advantage using photons},\ }\href {https://doi.org/10.1126/science.abe8770} {\bibfield  {journal} {\bibinfo  {journal} {Science}\ }\textbf {\bibinfo {volume} {370}},\ \bibinfo {pages} {1460} (\bibinfo {year} {2020})},\ \Eprint {https://arxiv.org/abs/https://www.science.org/doi/pdf/10.1126/science.abe8770} {https://www.science.org/doi/pdf/10.1126/science.abe8770} \BibitemShut {NoStop}%
\bibitem [{\citenamefont {Madsen}\ \emph {et~al.}(2022)\citenamefont {Madsen}, \citenamefont {Laudenbach}, \citenamefont {Askarani}, \citenamefont {Rortais}, \citenamefont {Vincent}, \citenamefont {Bulmer}, \citenamefont {Miatto}, \citenamefont {Neuhaus}, \citenamefont {Helt}, \citenamefont {Collins}, \citenamefont {Lita}, \citenamefont {Gerrits}, \citenamefont {Nam}, \citenamefont {Vaidya}, \citenamefont {Menotti}, \citenamefont {Dhand}, \citenamefont {Vernon}, \citenamefont {Quesada},\ and\ \citenamefont {Lavoie}}]{Squeeze_QComp_Madsen}%
  \BibitemOpen
  \bibfield  {author} {\bibinfo {author} {\bibfnamefont {L.~S.}\ \bibnamefont {Madsen}}, \bibinfo {author} {\bibfnamefont {F.}~\bibnamefont {Laudenbach}}, \bibinfo {author} {\bibfnamefont {M.~F.}\ \bibnamefont {Askarani}}, \bibinfo {author} {\bibfnamefont {F.}~\bibnamefont {Rortais}}, \bibinfo {author} {\bibfnamefont {T.}~\bibnamefont {Vincent}}, \bibinfo {author} {\bibfnamefont {J.~F.~F.}\ \bibnamefont {Bulmer}}, \bibinfo {author} {\bibfnamefont {F.~M.}\ \bibnamefont {Miatto}}, \bibinfo {author} {\bibfnamefont {L.}~\bibnamefont {Neuhaus}}, \bibinfo {author} {\bibfnamefont {L.~G.}\ \bibnamefont {Helt}}, \bibinfo {author} {\bibfnamefont {M.~J.}\ \bibnamefont {Collins}}, \bibinfo {author} {\bibfnamefont {A.~E.}\ \bibnamefont {Lita}}, \bibinfo {author} {\bibfnamefont {T.}~\bibnamefont {Gerrits}}, \bibinfo {author} {\bibfnamefont {S.~W.}\ \bibnamefont {Nam}}, \bibinfo {author} {\bibfnamefont {V.~D.}\ \bibnamefont {Vaidya}}, \bibinfo {author} {\bibfnamefont {M.}~\bibnamefont {Menotti}}, \bibinfo {author}
  {\bibfnamefont {I.}~\bibnamefont {Dhand}}, \bibinfo {author} {\bibfnamefont {Z.}~\bibnamefont {Vernon}}, \bibinfo {author} {\bibfnamefont {N.}~\bibnamefont {Quesada}},\ and\ \bibinfo {author} {\bibfnamefont {J.}~\bibnamefont {Lavoie}},\ }\bibfield  {title} {\bibinfo {title} {Quantum computational advantage with a programmable photonic processor},\ }\href {https://doi.org/10.1038/s41586-022-04725-x} {\bibfield  {journal} {\bibinfo  {journal} {Nature}\ }\textbf {\bibinfo {volume} {606}},\ \bibinfo {pages} {75} (\bibinfo {year} {2022})}\BibitemShut {NoStop}%
\bibitem [{\citenamefont {Dmytruk}\ and\ \citenamefont {Schir\'o}(2021)}]{dmytruk2021gauge}%
  \BibitemOpen
  \bibfield  {author} {\bibinfo {author} {\bibfnamefont {O.}~\bibnamefont {Dmytruk}}\ and\ \bibinfo {author} {\bibfnamefont {M.}~\bibnamefont {Schir\'o}},\ }\bibfield  {title} {\bibinfo {title} {Gauge fixing for strongly correlated electrons coupled to quantum light},\ }\href {https://doi.org/10.1103/PhysRevB.103.075131} {\bibfield  {journal} {\bibinfo  {journal} {Phys. Rev. B}\ }\textbf {\bibinfo {volume} {103}},\ \bibinfo {pages} {075131} (\bibinfo {year} {2021})}\BibitemShut {NoStop}%
\bibitem [{\citenamefont {Kozin}\ \emph {et~al.}(2025)\citenamefont {Kozin}, \citenamefont {Thingstad}, \citenamefont {Loss},\ and\ \citenamefont {Klinovaja}}]{Kozin2025}%
  \BibitemOpen
  \bibfield  {author} {\bibinfo {author} {\bibfnamefont {V.~K.}\ \bibnamefont {Kozin}}, \bibinfo {author} {\bibfnamefont {E.}~\bibnamefont {Thingstad}}, \bibinfo {author} {\bibfnamefont {D.}~\bibnamefont {Loss}},\ and\ \bibinfo {author} {\bibfnamefont {J.}~\bibnamefont {Klinovaja}},\ }\bibfield  {title} {\bibinfo {title} {Cavity-enhanced superconductivity via band engineering},\ }\href {https://doi.org/10.1103/PhysRevB.111.035410} {\bibfield  {journal} {\bibinfo  {journal} {Phys. Rev. B}\ }\textbf {\bibinfo {volume} {111}},\ \bibinfo {pages} {035410} (\bibinfo {year} {2025})}\BibitemShut {NoStop}%
\bibitem [{\citenamefont {Dmytruk}\ \emph {et~al.}(2015)\citenamefont {Dmytruk}, \citenamefont {Trif},\ and\ \citenamefont {Simon}}]{dmytruk2015cavity}%
  \BibitemOpen
  \bibfield  {author} {\bibinfo {author} {\bibfnamefont {O.}~\bibnamefont {Dmytruk}}, \bibinfo {author} {\bibfnamefont {M.}~\bibnamefont {Trif}},\ and\ \bibinfo {author} {\bibfnamefont {P.}~\bibnamefont {Simon}},\ }\bibfield  {title} {\bibinfo {title} {Cavity quantum electrodynamics with mesoscopic topological superconductors},\ }\href {https://doi.org/10.1103/PhysRevB.92.245432} {\bibfield  {journal} {\bibinfo  {journal} {Phys. Rev. B}\ }\textbf {\bibinfo {volume} {92}},\ \bibinfo {pages} {245432} (\bibinfo {year} {2015})}\BibitemShut {NoStop}%
\bibitem [{\citenamefont {P{\'e}rez-Gonz{\'a}lez}\ \emph {et~al.}(2022)\citenamefont {P{\'e}rez-Gonz{\'a}lez}, \citenamefont {G{\'o}mez-Le{\'o}n},\ and\ \citenamefont {Platero}}]{perez2022topology}%
  \BibitemOpen
  \bibfield  {author} {\bibinfo {author} {\bibfnamefont {B.}~\bibnamefont {P{\'e}rez-Gonz{\'a}lez}}, \bibinfo {author} {\bibfnamefont {{\'A}.}~\bibnamefont {G{\'o}mez-Le{\'o}n}},\ and\ \bibinfo {author} {\bibfnamefont {G.}~\bibnamefont {Platero}},\ }\bibfield  {title} {\bibinfo {title} {Topology detection in cavity {QED}},\ }\href {https://pubs.rsc.org/en/content/articlehtml/2022/cp/d2cp01806c} {\bibfield  {journal} {\bibinfo  {journal} {Physical Chemistry Chemical Physics}\ }\textbf {\bibinfo {volume} {24}},\ \bibinfo {pages} {15860} (\bibinfo {year} {2022})}\BibitemShut {NoStop}%
\bibitem [{\citenamefont {Hegde}\ \emph {et~al.}(2015)\citenamefont {Hegde}, \citenamefont {Shivamoggi}, \citenamefont {Vishveshwara},\ and\ \citenamefont {Sen}}]{hegde2015quench}%
  \BibitemOpen
  \bibfield  {author} {\bibinfo {author} {\bibfnamefont {S.}~\bibnamefont {Hegde}}, \bibinfo {author} {\bibfnamefont {V.}~\bibnamefont {Shivamoggi}}, \bibinfo {author} {\bibfnamefont {S.}~\bibnamefont {Vishveshwara}},\ and\ \bibinfo {author} {\bibfnamefont {D.}~\bibnamefont {Sen}},\ }\bibfield  {title} {\bibinfo {title} {Quench dynamics and parity blocking in {M}ajorana wires},\ }\href {https://doi.org/10.1088/1367-2630/17/5/053036} {\bibfield  {journal} {\bibinfo  {journal} {New Journal of Physics}\ }\textbf {\bibinfo {volume} {17}},\ \bibinfo {pages} {053036} (\bibinfo {year} {2015})}\BibitemShut {NoStop}%
\bibitem [{\citenamefont {Hegde}\ and\ \citenamefont {Vishveshwara}(2016)}]{hegde2016majorana}%
  \BibitemOpen
  \bibfield  {author} {\bibinfo {author} {\bibfnamefont {S.~S.}\ \bibnamefont {Hegde}}\ and\ \bibinfo {author} {\bibfnamefont {S.}~\bibnamefont {Vishveshwara}},\ }\bibfield  {title} {\bibinfo {title} {Majorana wave-function oscillations, fermion parity switches, and disorder in {K}itaev chains},\ }\href {https://doi.org/10.1103/PhysRevB.94.115166} {\bibfield  {journal} {\bibinfo  {journal} {Phys. Rev. B}\ }\textbf {\bibinfo {volume} {94}},\ \bibinfo {pages} {115166} (\bibinfo {year} {2016})}\BibitemShut {NoStop}%
\bibitem [{\citenamefont {Ran{\v c}i{\'c}}(2022)}]{rancic2022}%
  \BibitemOpen
  \bibfield  {author} {\bibinfo {author} {\bibfnamefont {M.~J.}\ \bibnamefont {Ran{\v c}i{\'c}}},\ }\bibfield  {title} {\bibinfo {title} {Exactly solving the {K}itaev chain and generating {M}ajorana-zero-modes out of noisy qubits},\ }\href {https://doi.org/10.1038/s41598-022-24341-z} {\bibfield  {journal} {\bibinfo  {journal} {Scientific Reports}\ }\textbf {\bibinfo {volume} {12}},\ \bibinfo {pages} {19882} (\bibinfo {year} {2022})}\BibitemShut {NoStop}%
\bibitem [{\citenamefont {Doicin}\ \emph {et~al.}(2025)\citenamefont {Doicin}, \citenamefont {Armour},\ and\ \citenamefont {Tufarelli}}]{LevRepulsion2025}%
  \BibitemOpen
  \bibfield  {author} {\bibinfo {author} {\bibfnamefont {T.}~\bibnamefont {Doicin}}, \bibinfo {author} {\bibfnamefont {A.}~\bibnamefont {Armour}},\ and\ \bibinfo {author} {\bibfnamefont {T.}~\bibnamefont {Tufarelli}},\ }\bibfield  {title} {\bibinfo {title} {Multilevel quantum {R}abi models},\ }\href {https://doi.org/10.1088/1751-8121/aded50} {\bibfield  {journal} {\bibinfo  {journal} {Journal of Physics A: Mathematical and Theoretical}\ }\textbf {\bibinfo {volume} {58}} (\bibinfo {year} {2025})}\BibitemShut {NoStop}%
\bibitem [{\citenamefont {Plumb}\ \emph {et~al.}(2016)\citenamefont {Plumb}, \citenamefont {Hwang}, \citenamefont {Qiu}, \citenamefont {Harriger}, \citenamefont {Granroth}, \citenamefont {Kolesnikov}, \citenamefont {Shu}, \citenamefont {Chou}, \citenamefont {R{\"u}egg}, \citenamefont {Kim},\ and\ \citenamefont {Kim}}]{LevelRepulsion2016}%
  \BibitemOpen
  \bibfield  {author} {\bibinfo {author} {\bibfnamefont {K.~W.}\ \bibnamefont {Plumb}}, \bibinfo {author} {\bibfnamefont {K.}~\bibnamefont {Hwang}}, \bibinfo {author} {\bibfnamefont {Y.}~\bibnamefont {Qiu}}, \bibinfo {author} {\bibfnamefont {L.~W.}\ \bibnamefont {Harriger}}, \bibinfo {author} {\bibfnamefont {G.~E.}\ \bibnamefont {Granroth}}, \bibinfo {author} {\bibfnamefont {A.~I.}\ \bibnamefont {Kolesnikov}}, \bibinfo {author} {\bibfnamefont {G.~J.}\ \bibnamefont {Shu}}, \bibinfo {author} {\bibfnamefont {F.~C.}\ \bibnamefont {Chou}}, \bibinfo {author} {\bibfnamefont {C.}~\bibnamefont {R{\"u}egg}}, \bibinfo {author} {\bibfnamefont {Y.~B.}\ \bibnamefont {Kim}},\ and\ \bibinfo {author} {\bibfnamefont {Y.-J.}\ \bibnamefont {Kim}},\ }\bibfield  {title} {\bibinfo {title} {Quasiparticle-continuum level repulsion in a quantum magnet},\ }\href {https://doi.org/10.1038/nphys3566} {\bibfield  {journal} {\bibinfo  {journal} {Nature Physics}\ }\textbf {\bibinfo {volume} {12}},\ \bibinfo {pages} {224} (\bibinfo {year}
  {2016})}\BibitemShut {NoStop}%
\bibitem [{\citenamefont {Reithmaier}\ \emph {et~al.}(2004)\citenamefont {Reithmaier}, \citenamefont {Sek}, \citenamefont {L{\"o}ffler}, \citenamefont {Hofmann}, \citenamefont {Kuhn}, \citenamefont {Reitzenstein}, \citenamefont {Keldysh}, \citenamefont {Kulakovskii}, \citenamefont {Reinecke},\ and\ \citenamefont {Forchel}}]{Anticross_Reithmaier}%
  \BibitemOpen
  \bibfield  {author} {\bibinfo {author} {\bibfnamefont {J.~P.}\ \bibnamefont {Reithmaier}}, \bibinfo {author} {\bibfnamefont {G.}~\bibnamefont {Sek}}, \bibinfo {author} {\bibfnamefont {A.}~\bibnamefont {L{\"o}ffler}}, \bibinfo {author} {\bibfnamefont {C.}~\bibnamefont {Hofmann}}, \bibinfo {author} {\bibfnamefont {S.}~\bibnamefont {Kuhn}}, \bibinfo {author} {\bibfnamefont {S.}~\bibnamefont {Reitzenstein}}, \bibinfo {author} {\bibfnamefont {L.~V.}\ \bibnamefont {Keldysh}}, \bibinfo {author} {\bibfnamefont {V.~D.}\ \bibnamefont {Kulakovskii}}, \bibinfo {author} {\bibfnamefont {T.~L.}\ \bibnamefont {Reinecke}},\ and\ \bibinfo {author} {\bibfnamefont {A.}~\bibnamefont {Forchel}},\ }\bibfield  {title} {\bibinfo {title} {Strong coupling in a single quantum dot--semiconductor microcavity system},\ }\href {https://doi.org/10.1038/nature02969} {\bibfield  {journal} {\bibinfo  {journal} {Nature}\ }\textbf {\bibinfo {volume} {432}},\ \bibinfo {pages} {197} (\bibinfo {year} {2004})}\BibitemShut {NoStop}%
\bibitem [{\citenamefont {Yoshie}\ \emph {et~al.}(2004)\citenamefont {Yoshie}, \citenamefont {Scherer}, \citenamefont {Hendrickson}, \citenamefont {Khitrova}, \citenamefont {Gibbs}, \citenamefont {Rupper}, \citenamefont {Ell}, \citenamefont {Shchekin},\ and\ \citenamefont {Deppe}}]{Anticross_Yoshie}%
  \BibitemOpen
  \bibfield  {author} {\bibinfo {author} {\bibfnamefont {T.}~\bibnamefont {Yoshie}}, \bibinfo {author} {\bibfnamefont {A.}~\bibnamefont {Scherer}}, \bibinfo {author} {\bibfnamefont {J.}~\bibnamefont {Hendrickson}}, \bibinfo {author} {\bibfnamefont {G.}~\bibnamefont {Khitrova}}, \bibinfo {author} {\bibfnamefont {H.~M.}\ \bibnamefont {Gibbs}}, \bibinfo {author} {\bibfnamefont {G.}~\bibnamefont {Rupper}}, \bibinfo {author} {\bibfnamefont {C.}~\bibnamefont {Ell}}, \bibinfo {author} {\bibfnamefont {O.~B.}\ \bibnamefont {Shchekin}},\ and\ \bibinfo {author} {\bibfnamefont {D.~G.}\ \bibnamefont {Deppe}},\ }\bibfield  {title} {\bibinfo {title} {Vacuum {R}abi splitting with a single quantum dot in a photonic crystal nanocavity},\ }\href {https://doi.org/10.1038/nature03119} {\bibfield  {journal} {\bibinfo  {journal} {Nature}\ }\textbf {\bibinfo {volume} {432}},\ \bibinfo {pages} {200} (\bibinfo {year} {2004})}\BibitemShut {NoStop}%
\bibitem [{\citenamefont {Peter}\ \emph {et~al.}(2005)\citenamefont {Peter}, \citenamefont {Senellart}, \citenamefont {Martrou}, \citenamefont {Lema\^{\i}tre}, \citenamefont {Hours}, \citenamefont {G\'erard},\ and\ \citenamefont {Bloch}}]{Anticross_Peter}%
  \BibitemOpen
  \bibfield  {author} {\bibinfo {author} {\bibfnamefont {E.}~\bibnamefont {Peter}}, \bibinfo {author} {\bibfnamefont {P.}~\bibnamefont {Senellart}}, \bibinfo {author} {\bibfnamefont {D.}~\bibnamefont {Martrou}}, \bibinfo {author} {\bibfnamefont {A.}~\bibnamefont {Lema\^{\i}tre}}, \bibinfo {author} {\bibfnamefont {J.}~\bibnamefont {Hours}}, \bibinfo {author} {\bibfnamefont {J.~M.}\ \bibnamefont {G\'erard}},\ and\ \bibinfo {author} {\bibfnamefont {J.}~\bibnamefont {Bloch}},\ }\bibfield  {title} {\bibinfo {title} {Exciton-photon strong-coupling regime for a single quantum dot embedded in a microcavity},\ }\href {https://doi.org/10.1103/PhysRevLett.95.067401} {\bibfield  {journal} {\bibinfo  {journal} {Phys. Rev. Lett.}\ }\textbf {\bibinfo {volume} {95}},\ \bibinfo {pages} {067401} (\bibinfo {year} {2005})}\BibitemShut {NoStop}%
\bibitem [{\citenamefont {Gergs}\ \emph {et~al.}(2016)\citenamefont {Gergs}, \citenamefont {Fritz},\ and\ \citenamefont {Schuricht}}]{PhysRevB.93.075129}%
  \BibitemOpen
  \bibfield  {author} {\bibinfo {author} {\bibfnamefont {N.~M.}\ \bibnamefont {Gergs}}, \bibinfo {author} {\bibfnamefont {L.}~\bibnamefont {Fritz}},\ and\ \bibinfo {author} {\bibfnamefont {D.}~\bibnamefont {Schuricht}},\ }\bibfield  {title} {\bibinfo {title} {Topological order in the {K}itaev/{M}ajorana chain in the presence of disorder and interactions},\ }\href {https://doi.org/10.1103/PhysRevB.93.075129} {\bibfield  {journal} {\bibinfo  {journal} {Phys. Rev. B}\ }\textbf {\bibinfo {volume} {93}},\ \bibinfo {pages} {075129} (\bibinfo {year} {2016})}\BibitemShut {NoStop}%
\bibitem [{\citenamefont {del Pozo}\ \emph {et~al.}(2025)\citenamefont {del Pozo}, \citenamefont {Herviou}, \citenamefont {Dmytruk},\ and\ \citenamefont {Le~Hur}}]{PhysRevB.111.075170}%
  \BibitemOpen
  \bibfield  {author} {\bibinfo {author} {\bibfnamefont {F.}~\bibnamefont {del Pozo}}, \bibinfo {author} {\bibfnamefont {L.}~\bibnamefont {Herviou}}, \bibinfo {author} {\bibfnamefont {O.}~\bibnamefont {Dmytruk}},\ and\ \bibinfo {author} {\bibfnamefont {K.}~\bibnamefont {Le~Hur}},\ }\bibfield  {title} {\bibinfo {title} {Model for topological $p$-wave superconducting wires with disorder and interactions},\ }\href {https://doi.org/10.1103/PhysRevB.111.075170} {\bibfield  {journal} {\bibinfo  {journal} {Phys. Rev. B}\ }\textbf {\bibinfo {volume} {111}},\ \bibinfo {pages} {075170} (\bibinfo {year} {2025})}\BibitemShut {NoStop}%
\bibitem [{\citenamefont {Pekerten}\ \emph {et~al.}(2019)\citenamefont {Pekerten}, \citenamefont {Bozkurt},\ and\ \citenamefont {Adagideli}}]{Majobilliards}%
  \BibitemOpen
  \bibfield  {author} {\bibinfo {author} {\bibfnamefont {B.}~\bibnamefont {Pekerten}}, \bibinfo {author} {\bibfnamefont {A.~M.}\ \bibnamefont {Bozkurt}},\ and\ \bibinfo {author} {\bibfnamefont {{\.I}.}~\bibnamefont {Adagideli}},\ }\bibfield  {title} {\bibinfo {title} {Fermion parity switches of the ground state of {M}ajorana billiards},\ }\href {https://doi.org/10.1103/PhysRevB.100.235455} {\bibfield  {journal} {\bibinfo  {journal} {Physical Review B}\ }\textbf {\bibinfo {volume} {100}},\ \bibinfo {pages} {235455} (\bibinfo {year} {2019})}\BibitemShut {NoStop}%
\bibitem [{\citenamefont {Bertet}\ \emph {et~al.}(2002)\citenamefont {Bertet}, \citenamefont {Auffeves}, \citenamefont {Maioli}, \citenamefont {Osnaghi}, \citenamefont {Meunier}, \citenamefont {Brune}, \citenamefont {Raimond},\ and\ \citenamefont {Haroche}}]{CavExp_Bertet}%
  \BibitemOpen
  \bibfield  {author} {\bibinfo {author} {\bibfnamefont {P.}~\bibnamefont {Bertet}}, \bibinfo {author} {\bibfnamefont {A.}~\bibnamefont {Auffeves}}, \bibinfo {author} {\bibfnamefont {P.}~\bibnamefont {Maioli}}, \bibinfo {author} {\bibfnamefont {S.}~\bibnamefont {Osnaghi}}, \bibinfo {author} {\bibfnamefont {T.}~\bibnamefont {Meunier}}, \bibinfo {author} {\bibfnamefont {M.}~\bibnamefont {Brune}}, \bibinfo {author} {\bibfnamefont {J.~M.}\ \bibnamefont {Raimond}},\ and\ \bibinfo {author} {\bibfnamefont {S.}~\bibnamefont {Haroche}},\ }\bibfield  {title} {\bibinfo {title} {Direct measurement of the {W}igner function of a one-photon fock state in a cavity},\ }\href {https://doi.org/10.1103/PhysRevLett.89.200402} {\bibfield  {journal} {\bibinfo  {journal} {Phys. Rev. Lett.}\ }\textbf {\bibinfo {volume} {89}},\ \bibinfo {pages} {200402} (\bibinfo {year} {2002})}\BibitemShut {NoStop}%
\bibitem [{\citenamefont {Schuster}\ \emph {et~al.}(2007)\citenamefont {Schuster}, \citenamefont {Houck}, \citenamefont {Schreier}, \citenamefont {Wallraff}, \citenamefont {Gambetta}, \citenamefont {Blais}, \citenamefont {Frunzio}, \citenamefont {Majer}, \citenamefont {Johnson}, \citenamefont {Devoret}, \citenamefont {Girvin},\ and\ \citenamefont {Schoelkopf}}]{CavExp_Schuster}%
  \BibitemOpen
  \bibfield  {author} {\bibinfo {author} {\bibfnamefont {D.~I.}\ \bibnamefont {Schuster}}, \bibinfo {author} {\bibfnamefont {A.~A.}\ \bibnamefont {Houck}}, \bibinfo {author} {\bibfnamefont {J.~A.}\ \bibnamefont {Schreier}}, \bibinfo {author} {\bibfnamefont {A.}~\bibnamefont {Wallraff}}, \bibinfo {author} {\bibfnamefont {J.~M.}\ \bibnamefont {Gambetta}}, \bibinfo {author} {\bibfnamefont {A.}~\bibnamefont {Blais}}, \bibinfo {author} {\bibfnamefont {L.}~\bibnamefont {Frunzio}}, \bibinfo {author} {\bibfnamefont {J.}~\bibnamefont {Majer}}, \bibinfo {author} {\bibfnamefont {B.}~\bibnamefont {Johnson}}, \bibinfo {author} {\bibfnamefont {M.~H.}\ \bibnamefont {Devoret}}, \bibinfo {author} {\bibfnamefont {S.~M.}\ \bibnamefont {Girvin}},\ and\ \bibinfo {author} {\bibfnamefont {R.~J.}\ \bibnamefont {Schoelkopf}},\ }\bibfield  {title} {\bibinfo {title} {Resolving photon number states in a superconducting circuit},\ }\href {https://doi.org/10.1038/nature05461} {\bibfield  {journal} {\bibinfo  {journal} {Nature}\ }\textbf
  {\bibinfo {volume} {445}},\ \bibinfo {pages} {515} (\bibinfo {year} {2007})}\BibitemShut {NoStop}%
\bibitem [{\citenamefont {Andolina}\ \emph {et~al.}(2022)\citenamefont {Andolina}, \citenamefont {Pellegrino}, \citenamefont {Mercurio}, \citenamefont {Di~Stefano}, \citenamefont {Polini},\ and\ \citenamefont {Savasta}}]{Andolina_azero}%
  \BibitemOpen
  \bibfield  {author} {\bibinfo {author} {\bibfnamefont {G.~M.}\ \bibnamefont {Andolina}}, \bibinfo {author} {\bibfnamefont {F.~M.~D.}\ \bibnamefont {Pellegrino}}, \bibinfo {author} {\bibfnamefont {A.}~\bibnamefont {Mercurio}}, \bibinfo {author} {\bibfnamefont {O.}~\bibnamefont {Di~Stefano}}, \bibinfo {author} {\bibfnamefont {M.}~\bibnamefont {Polini}},\ and\ \bibinfo {author} {\bibfnamefont {S.}~\bibnamefont {Savasta}},\ }\bibfield  {title} {\bibinfo {title} {A non-perturbative no-go theorem for photon condensation in approximate models},\ }\href {https://doi.org/10.1140/epjp/s13360-022-03571-0} {\bibfield  {journal} {\bibinfo  {journal} {The European Physical Journal Plus}\ }\textbf {\bibinfo {volume} {137}},\ \bibinfo {pages} {1348} (\bibinfo {year} {2022})}\BibitemShut {NoStop}%
\bibitem [{\citenamefont {Andolina}\ \emph {et~al.}(2019)\citenamefont {Andolina}, \citenamefont {Pellegrino}, \citenamefont {Giovannetti}, \citenamefont {MacDonald},\ and\ \citenamefont {Polini}}]{Andolina_no-go_theorem}%
  \BibitemOpen
  \bibfield  {author} {\bibinfo {author} {\bibfnamefont {G.~M.}\ \bibnamefont {Andolina}}, \bibinfo {author} {\bibfnamefont {F.~M.~D.}\ \bibnamefont {Pellegrino}}, \bibinfo {author} {\bibfnamefont {V.}~\bibnamefont {Giovannetti}}, \bibinfo {author} {\bibfnamefont {A.~H.}\ \bibnamefont {MacDonald}},\ and\ \bibinfo {author} {\bibfnamefont {M.}~\bibnamefont {Polini}},\ }\bibfield  {title} {\bibinfo {title} {Cavity quantum electrodynamics of strongly correlated electron systems: {A} no-go theorem for photon condensation},\ }\href {https://doi.org/10.1103/PhysRevB.100.121109} {\bibfield  {journal} {\bibinfo  {journal} {Phys. Rev. B}\ }\textbf {\bibinfo {volume} {100}},\ \bibinfo {pages} {121109} (\bibinfo {year} {2019})}\BibitemShut {NoStop}%
\bibitem [{\citenamefont {Mikami}\ \emph {et~al.}(2016)\citenamefont {Mikami}, \citenamefont {Kitamura}, \citenamefont {Yasuda}, \citenamefont {Tsuji}, \citenamefont {Oka},\ and\ \citenamefont {Aoki}}]{HighFrequencyExp}%
  \BibitemOpen
  \bibfield  {author} {\bibinfo {author} {\bibfnamefont {T.}~\bibnamefont {Mikami}}, \bibinfo {author} {\bibfnamefont {S.}~\bibnamefont {Kitamura}}, \bibinfo {author} {\bibfnamefont {K.}~\bibnamefont {Yasuda}}, \bibinfo {author} {\bibfnamefont {N.}~\bibnamefont {Tsuji}}, \bibinfo {author} {\bibfnamefont {T.}~\bibnamefont {Oka}},\ and\ \bibinfo {author} {\bibfnamefont {H.}~\bibnamefont {Aoki}},\ }\bibfield  {title} {\bibinfo {title} {Brillouin-{W}igner theory for high-frequency expansion in periodically driven systems: {A}pplication to {F}loquet topological insulators},\ }\href {https://doi.org/10.1103/PhysRevB.93.144307} {\bibfield  {journal} {\bibinfo  {journal} {Phys. Rev. B}\ }\textbf {\bibinfo {volume} {93}},\ \bibinfo {pages} {144307} (\bibinfo {year} {2016})}\BibitemShut {NoStop}%
\bibitem [{\citenamefont {W\'odkiewicz}\ \emph {et~al.}(1987)\citenamefont {W\'odkiewicz}, \citenamefont {Knight}, \citenamefont {Buckle},\ and\ \citenamefont {Barnett}}]{squeeze_Wodkiewicz}%
  \BibitemOpen
  \bibfield  {author} {\bibinfo {author} {\bibfnamefont {K.}~\bibnamefont {W\'odkiewicz}}, \bibinfo {author} {\bibfnamefont {P.~L.}\ \bibnamefont {Knight}}, \bibinfo {author} {\bibfnamefont {S.~J.}\ \bibnamefont {Buckle}},\ and\ \bibinfo {author} {\bibfnamefont {S.~M.}\ \bibnamefont {Barnett}},\ }\bibfield  {title} {\bibinfo {title} {Squeezing and superposition states},\ }\href {https://doi.org/10.1103/PhysRevA.35.2567} {\bibfield  {journal} {\bibinfo  {journal} {Phys. Rev. A}\ }\textbf {\bibinfo {volume} {35}},\ \bibinfo {pages} {2567} (\bibinfo {year} {1987})}\BibitemShut {NoStop}%
\end{thebibliography}%

\end{document}